\documentclass{aa}  

\usepackage[pdfpagelabels=false]{hyperref}	
\hypersetup{colorlinks=true,linkcolor=blue,citecolor=blue,filecolor=blue,urlcolor=blue,}
\usepackage{natbib}
\usepackage{graphicx}
\usepackage{soul}
\usepackage[dvipsnames]{xcolor}
\usepackage{txfonts}
\usepackage{lipsum}
\usepackage{subcaption}  
\usepackage{lscape}             
\usepackage{placeins}           
                                
\begin{document}

   \title{Reconstructing chemical enrichment pathways in disc galaxies}
   \subtitle{A phylogenetic approach}

   \author{Brian Tapia-Contreras\inst{1, 2, 3}$^\dagger$ \and 
           Patricia B. Tissera\inst{1, 2, 3}\and
           Emanuel Sillero\inst{1, 2, 3}\and
           Paula Jofré\inst{3, 4}\and
           Keaghan Yaxley\inst{5}\and
           Xia Hua\inst{6}\and
           Robert M. Yates\inst{7}\and
           Álvaro Márquez S.\inst{3, 8}\and
           Theosamuele Signor\inst{4, 9}\and
           Payel Das\inst{10}\and
           Álvaro Rojas-Arriagada\inst{3, 11, 12, 13}\and
           Claudia Aguilera-Gómez\inst{1, 3}\and
           Francisco Jara-Ferreira\inst{1, 2, 3}\and
           Robert A. Foley\inst{14}
        }

   \institute{Instituto de Astrofísica, Pontificia Universidad Católica de Chile, Av. Vicuña Mackenna 4860, Santiago, Chile.\\
    $^\dagger$\email{brian.tapia@uc.cl} \and 
    Centro de Astro-ingeniería, Pontificia Universidad Católica de Chile, Av. Vicuña Mackenna 4860, Santiago, Chile.\and
    Millenium Nucleus ERIS\and
    Instituto de Estudios Astrofísicos, Facultad de Ingeniería y Ciencias, Univesidad Diego Portales, Santiago de Chile\and
    Macroevolution and Macroecology, Research School of Biology, Australian National University, Canberra, ACT 0200, Australia\and
    Mathematical Sciences Institute, Australian National University, Canberra ACT 0200 Australia\and
    Centre for Astrophysics Research, University of Hertfordshire, Hatfield AL10 9AB, UK\and
    Department of Mathematical Engineering, Universidad de Chile\and
    Inria Chile Research Center, Av. Apoquindo 2827, piso 12, Las Condes, Santiago, Chile\and
    School of Mathematics and Physics, University of Surrey, Guildford, Surrey, GU2 7XH, UK\and
    Departamento de Física, Universidad de Santiago de Chile, Av. Victor Jara 3659, Santiago, Chile\and
    Millenium Institute of Astrophysics (MAS), Av. Vicuña Mackenna 4860, 82-0436 Macul, Santiago, Chile\and
    Center for Interdisciplinary Research in Astrophysics and Space Exploration (CIRAS), Universidad de Santiago de Chile, Santiago, Chile\and
    Leverhulme Centre for Human Evolutionary Studies, Department for Anthropology and Archaeology, University of Cambridge, Fitzwilliam Street., Cambridge CB2 1QH, UK
    }

   \date{Received September 30, 20XX}


  \abstract
   { Phylogenetic methods, traditionally used in biology to trace the evolutionary relationships among species, are emerging as a powerful framework to reconstruct evolutionary processes in galaxies from chemical information.
   }
   {We apply galactic phylogenetics to study the chemical evolution of stellar populations in distinct regions of a simulated disc galaxy, assessing its capability to unveil assembly histories.} 
   {We used a high-resolution simulation that followed the chemical enrichment of an isolated disc galaxy by different stellar progenitors. We tracked gas particles as they turned into stars and inherited their parent gas chemical composition. Target particles were selected to store the chemical history of each chemical element considered in the simulation. Two regions were analysed: an inner ring, influenced by early bar-driven inflows, and an outer ring, shaped by spiral arms. We built phylogenetic trees for stellar populations in each region and quantified their structure using the Corrected Colless index, a standard metric of tree balance used in biology.}
   {The inner ring tree reveals a compact clade of old stars enriched by rapid Type II supernova (SNII) feedback, followed by a hierarchical sequence with increasing Type Ia supernova (SNIa) and asymptotic giant branch (AGB) contributions. In contrast, the outer ring exhibits more symmetric, caterpillar-like trees with smoother abundance gradients, consistent with more prolonged star formation and efficient local mixing. Chemical enrichment rates corroborate these trends, showing fast early enrichment in the inner ring and gradual, spatially extended enrichment in the outer disc. The structural indices differ significantly between the two regions and converge robustly even for modest stellar samples (N$_{\rm SSP} = 100)$.}
   {Galactic phylogenetics provides a novel and complementary tool to decode the fossil record of galaxies. It captures distinct chemical pathways across galactic environments and offers quantitative metrics for comparing assembly histories purely from chemical abundances.}


   \maketitle
   \nolinenumbers

\section{Introduction}

Galaxies are complex systems composed of stars, gas, dust, and dark matter, evolving through cosmic time while interacting with their surroundings. They grow by accreting gas and stars from the intergalactic medium and neighbouring systems, and the physical processes that shape their evolution, such as gas accretion, mergers, stellar feedback, and radial migration, operate across a wide range of spatial and temporal scales \citep[e.g.][]{whitefrenk1991}. These processes leave observable imprints on the properties of galaxies, particularly on the chemical composition of their stellar populations, which serves as a fossil record of their formation histories \citep{freeman}.

Chemical abundance patterns in stars of different ages are especially powerful tools to probe galaxy evolution, as they reflect the integrated action of nucleosynthesis, gas flows, and star formation over time. As baryons, originally in the form of primordial gas (mainly H and He, with traces of D and Li), are converted into stars, heavier elements are synthesised and returned to the interstellar medium (ISM) through supernovae and stellar winds. The ISM becomes progressively enriched in heavy elements, and new generations of stars inherit this chemical signature. This process, known as astration, creates a sequence of chemical patterns that can be linked to the physical mechanisms driving galaxy evolution \citep{tinsley1980}. However, disentangling the overlapping effects of different processes remains a significant challenge.

In this context, chemical evolution models have long served as essential tools to interpret the observed abundance patterns of stars. Detailed analytic and semi-analytic models describing how elements are produced and distributed by Type II and Type Ia supernovae (SNII and SNIa, respectively), as well as by asymptotic giant branch (AGB) stars, provide key insights into the timescales and efficiency of star formation in different galactic environments \citep{matteucci1986, molla1995}. These models have been compared to the Milky Way (MW) and other galaxies, revealing the connections between chemical signatures and star formation history \citep{Thomas1998, Nagashima2005, Calura2009, Arrigoni2010, Pipino2011, Yates2013}.

While analytic and semi-analytic models remain fundamental, they often lack the spatial and dynamical resolution needed to fully capture galaxy evolution. Hydrodynamical chemodynamical simulations address this limitation by self-consistently modelling both the gravitational and chemical evolution of galaxies \cite[e.g.][]{mosconi2001,lia2002}. These simulations follow gas dynamics, star formation, and feedback processes over cosmic time, incorporating element production and transport across the galaxy. They have become indispensable for linking observational data across cosmic time and validating theoretical models \citep[][]{rait1996, tissera2012,tissera2016,pedrosa2015,sillero2017,yates2021,bellardini2023,garcia2023}.  In particular, the implementation of chemical enrichment in cosmological simulations provides a powerful means to study the interplay between baryonic physics and chemical signatures within the evolving structure of galaxies \citep[e.g.][]{tissera2013, ma2017, derossi2017, hemler2020, fragkoudi2020, jara2024, tapia2025, tau2025, gonzalez2025}. Treating chemical evolution within the context of the non-linear formation and evolution of galaxies provides fundamental insights into how the chemical properties of baryons are linked to specific events in galaxy evolution, such as mergers \citep[e.g.][]{kewley2010, rupke2010, perez2011} or migration \citep{johnson2021}.

Simultaneously, advances in observational astronomy have revolutionised our ability to test these models. Multi-band photometric and spectroscopic surveys, together with astrometric data from the Gaia mission, have produced high-precision measurements of stellar positions, motions, and chemical abundances for millions of stars in the MW \citep[e.g.][]{jofre2017,queiroz2023,recioblanco2023}. This wealth of data enables the identification of chemically distinct populations and accreted components \citep[e.g.][]{helmi2020}, as well as the characterisation of radial gradients, age–metallicity relations, and dynamical substructures \citep[][and references therein]{deason2024}. As the MOONS, 4-MOST, Vera Rubin Observatory, and later the ELT, begin operations, the volume and quality of such data will keep growing exponentially.

The growing complexity of observational datasets calls for new approaches that can reconstruct the history of galaxy evolution from the chemical abundances of stellar populations, which preserve the imprint of the ISM from which they formed. A promising framework is provided by phylogenetic methods, originally developed in evolutionary biology to reconstruct the relationships among organisms based on heritable traits \citep{darwin1859}. A phylogeny is a graph whose branches depict both the pattern of relatedness among species, indicating which species are more closely related, and the evolutionary distance between species, which stores information on how much evolutionary change has accrued or how much time has passed since the species diverged from a shared common ancestor. The points on a phylogeny where branches converge are called nodes and represent the last common ancestor of the descending species, which themselves can be collectively referred to as a clade. The node representing the last common ancestor of all the species on the tree is referred to as the root of the tree. The amount of evolutionary distance between nodes is captured by the length of the branches separating them.

Under the assumption that chemical abundances behave analogously to DNA, which is transmitted  through descent with modification from one generation to the next, phylogenetic tools have been already applied to observed stellar populations  in the disc of the MW \citep{jofre2017,jackson2021, walsen2024} and to reconstruct the complex evolutionary history of Omega Centauri \citep[$\omega$Cen;][]{jofre2025}. This assumption is based on the fact that new chemical elements are injected into the ISM by the previous stellar generations, modifying the abundances of the gas clouds from which new stars are formed. The new stars inherit their chemical patterns and, hence, are used as tracers of chemical evolution.

Additionally, \citet{debrito2024} first tested this approach in numerical simulations, as  they offer a powerful framework for connecting the properties of the trees to specific events in the assembly history of galaxies.  \citet{debrito2024} showed that in an isolated disc galaxy, with neither interactions nor external gas infall, the phylogenetic tree is largely featureless, resembling a single sequence or a `caterpillar' \footnote{A hierarchical tree characterised by minimal branching and the absence of prominent clades, as expected when stars form from an ISM continuously enriched by previous stellar generations, without exchange of material with the surrounding medium.} tree. In such trees, stellar ages evolve monotonically from the root. Moreover, different regions within the same galaxy give rise to longer (shorter) trees in environments with higher (lower) star formation rates. When applied to observations in the case of the MW and $\omega$Cen, phylogenetic trees displayed distinct clades. Numerical simulations provide a crucial framework for linking the chemo-dynamical evolution of galaxies to the structure of their phylogenetic trees, thereby enabling a physically grounded interpretation of observed phylogenetic trees.

In this paper, we advance the understanding of galactic phylogenetics using numerical simulations. We apply galactic phylogenetics to a chemodynamical simulation of an isolated disc galaxy, dubbed the Origins simulation, tailored to study the chemical enrichment of different galactic regions. We store detailed information on the chemical enrichment history by explicitly tracking the production, transport, and final fate of individual chemical elements synthesised by SNII, SNIa, and AGB stars. This enables a deeper interpretation of the information encoded in the phylogenetic trees. We follow the chemical evolution of two sets of gas particles selected from an inner and outer ring of the disc, and trace their transformation into stars. By constructing phylogenetic trees purely based on the resulting stellar abundances, we investigate whether tree structure encodes differences in star formation and enrichment history between regions. Additional information, such as positions, velocities, and ages, is widely used in astronomy to assess the membership of a star in a given group using clustering algorithms. We emphasise that our approach relies on chemical information alone, while still providing evolutionary insight through the structure of the trees. This study takes a step further by assessing the potential of phylogenetic methods in a controlled chemo-dynamical framework, aiming to complement classical modelling and simulations with a novel approach to decoding the chemical memory of galaxies.

This paper is organised as follows. Sect.~\ref{sec:sim} describes the numerical experiment. Sect.~\ref{sec:history} discusses the enrichment histories of the selected target particles. Sect.~\ref{sec:phylo} revisits galactic phylogenetics, expands upon the global concepts outlined in the Introduction and applies the method to reconstruct chemical enrichment histories. In Sect.~\ref{sec:discussion}, the phylogenetic trees are interpreted using the numerical histories and compared through standard measures borrowed from biology. Finally, in the Conclusions, we summarise our main results.

\section{The Origins simulation}
\label{sec:sim}
For this work, we designed and performed a tailored simulation, dubbed Origins, representing  a Milky Way-mass  galaxy evolving in isolation. The initial condition (IC) of Origins adopts an exponential disc with a total mass of $\rm M_{\ast} = 1.2 \times 10^{10} M_{\odot}$ with a minimum gas mass resolution of $\bf \rm 1.2 \times 10^{5} M_{\odot}$. 
It contains an initial gas fraction of $f_{\rm gas} = 0.20$. The bulge component is represented with a Hernquist potential \citep{Hernquist1990}, while the dark matter halo is modelled using a NFW potential with $c=14$ \citep{Navarro1996, Navarro1997}.
The gravitational softening is 0.20 kpc for the gas and star particles, and 0.32 kpc for the dark matter ones. This simulation was run for about 3 Gyr with a temporal cadence of 5 Myr. The IC was run without allowing star formation to take place until the bar formed and stabilised. This was done to avoid a significant fraction of the gas near the central region falling into the bar. However, as explained in the following section, some gas particles in the nearby region did reach the very central region.

The simulation was performed by using an updated version of the Smoothed Particle Hydrodynamical (SPH) code  {\sc P-GADGET-3}, which includes a multiphase treatment for the ISM and the SN feedback models reported by \citet{scan05, scan06}. 
The chemical model was updated from that used in the CIELO project \citep[][]{rodriguez2022, cataldi2023, casanueva2024, gonzalez2025,tissera2025}. The current version includes metal-dependent cooling rates reported by \citet[][see Sillero et al. in prep.]{wiersma2009a}. For optically thin gas inionisation equilibrium, these rates were calculated on an element-by-element basis and consider the effects of photoionisation from a uniform, redshift-dependent ionising background \citep{haardt2001}. 

The initial gas composition is primordial, containing only hydrogen and helium with mass fractions $X = 0.76$ and $Y = 0.24$, respectively. The metallicity was set to $Z = 0$, where $Z$ represents the mass fraction of all elements heavier than hydrogen and helium. The gas is then transformed into stars if it satisfies density and temperature thresholds following the Kennicutt-Schmidt law \citep[e.g.][]{pedrosa2015, sillero2017}.  We adopted an initial  mass function  of \citet{salpeter1955},  with minimum and maximum mass cut-offs of  0.1 and 120 $\rm M_{\odot}$, respectively. Each gas particle represents a gas cloud with certain chemical abundances that are inherited by a star particle, representing a single stellar population (SSP). For simplicity, we refer to gas and star particles throughout the paper.  

\subsection{Chemical evolution model}
\label{sec:chemicalmodel}

The chemical enrichment was produced by SNII\, SNIa, and AGB stars.
We followed 22 isotopes: $^{1}$H, $^{4}$He, $^{12}$C, $^{13}$C, $^{14}$N, $^{15}$N, $^{16}$O, $^{20}$Ne, $^{22}$Ne, $^{23}$Na, $^{24}$Mg, $^{25}$Mg, $^{26}$Mg, $^{27}$Al, $^{28}$Si, $^{32}$S, $^{40}$Ca, $^{54}$Fe, and $^{56}$Fe.

For SNII, the production of new elements was calculated using the  nucleosynthesis models of \citet{Nomoto2013}. The lifetimes of massive stars that end as SNII were estimated by  using the metallicity-dependent scheme reported by \citet{rait1996}. 
For SNIa, we assumed the delay time distribution (DTD) scheme associated with a single degenerate progenitor scenario \citep[][see also \citet{Jimenez2015}]{mosconi2001, matteucci2021} for the binary system. The production of new elements by SNIa was taken from  \citet{Iwamoto1999}. Finally,  the isotopes ejected by stars in the AGB stage were taken from \citet{Karakas2010}, which are a function of metallicity and age. The AGB nucleosynthesis contributions are only injected into the cold gas phase, while the enriched material expelled by SNe is distributed 70\% in the cold phase and the remaining in the hot phase \citep[see][]{scan05}.

The energy injected by SNII and SNIa are equally distributed within the cold and hot environments of the corresponding SSP. The energy injected into the hot phase is instantaneously thermalised, while the energy deposited in the cold phase is reserved until it accumulates to a level comparable to the average energy present in neighbouring hot gas particles and therefore sufficient to promote the cold gas to a hot phase. 

\begin{figure*}[t]
\centering
\includegraphics[width=0.42\textwidth]{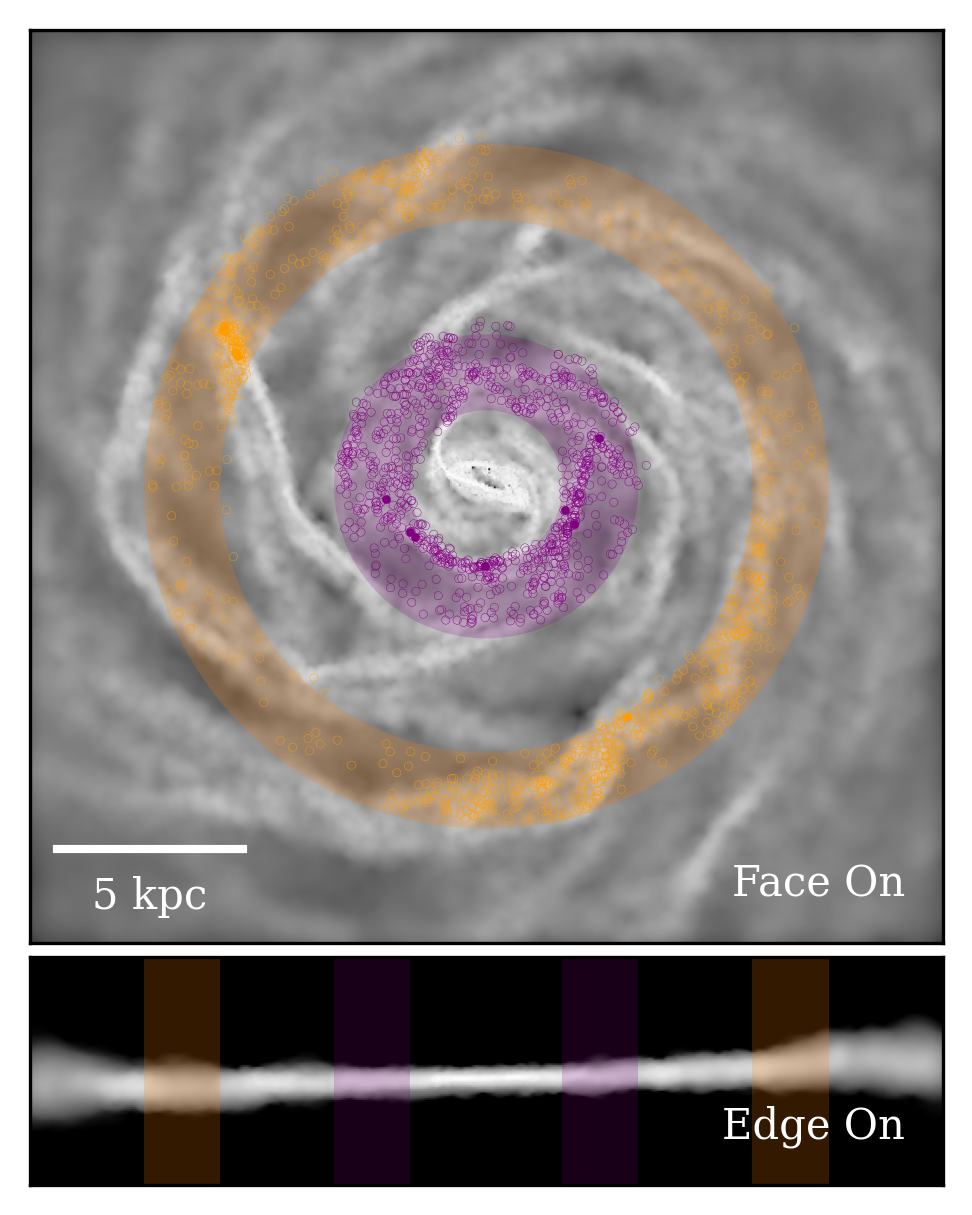}
\includegraphics[width=0.42\textwidth]{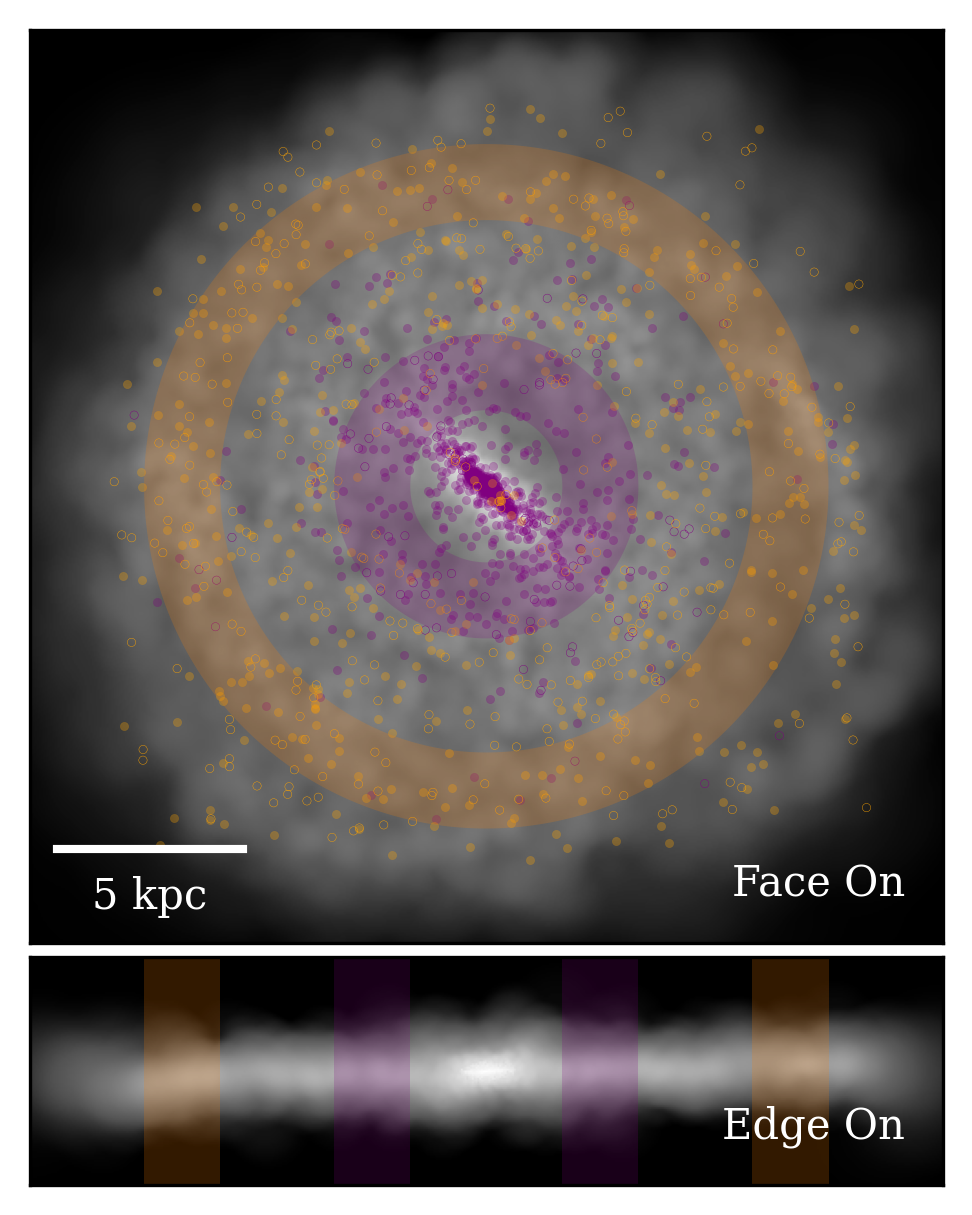}
\caption{Projected gas (left) and stellar (right) density distributions of the simulated galaxy. The inner ($r \in [2, 4]$ kpc) and outer ($r \in [7, 9]$ kpc) rings are indicated by the shaded purple and orange regions, respectively. The left panel shows the initial gas distribution and the right panel the final stellar distribution. Empty and filled circles depict target gas and stellar particles, respectively.
\label{fig:maps}}
\end{figure*}

The ejection and mixing of chemical elements is an important process that, in this simulation, was modelled following \citet{mosconi2001}. These authors proposed to weight the metal contribution to the neighbouring gas particles of a given donor star by the same kernel used by the SPH to estimate the hydrodynamical properties \citep[see also][]{wiersma2009a}. We note, however, that the smoothing lengths used to distribute the chemical elements were estimated on a particle-by-particle basis at each time that this process was needed. Hence, this produced different mixing of the chemical elements produced by SNII, SNIa, or AGB stars because they have different lifetimes and gas and stars follow distinct evolutionary paths. As a result, the differential mixing can imprint different features in the chemical patterns of the stellar populations \citep[see also][]{zhang2025}, which could be reflected in the structure of the phylogenetic trees.

To assess the capability of phylogenetic trees to decode the chemical evolution of the ISM using stellar populations, we stored the complete exchange of material between stellar and gas particles.
To follow the enrichment history of gas particles, we included a feature in the code to store the contributions of chemical elements received by a gas particle (hereafter referred to as the target) from surrounding stellar particles (hereafter referred to as the donors). This exchange of chemical elements was stored until the gas particle was converted into stars. Additionally, we tracked the chemical contributions produced by new star particles and injected into neighbouring gas particles throughout the integration period. In this way, we followed the enrichment history on a particle-by-particle basis, and identified the origin of each contribution produced by a donor stellar particle or received by a target gas particle.
It is worth noting that only the abundances of gas particles evolve as a result of stellar evolution. The chemical content of each stellar particle remains fixed, matching the chemical abundances of the progenitor gas particle at the time of its transformation into stars.
Given the large volume of data involved in tracking the full exchange of material between particles, we adopted a tracer-based approach by selecting a representative subset of gas particles at the beginning of the simulation, hereafter referred to as target particles, to follow their chemical enrichment history. As explain in the following section, we defined two regions of interest in the disc: one adjacent to the bulge, and another located in the disc component. We selected the target particles within each of the two regions.

\subsection{Target regions}\label{sec:sample_selection}

The target particles were selected at the initial time of the simulation, $t=0$. We defined two rings, each 2 kpc wide, and randomly selected 1000 gas particles in each of them. The so-called inner and outer rings were located  at 3 kpc and 8 kpc from the galactic centre. Because of the dynamical evolution, spiral arms and a bar were formed. These features  are efficient at redistributing angular momentum and hence, particles. In doing so, they contribute to mixing the gas and stars, sometimes displacing them from their host rings. We stress that the gas particles in both rings started with primordial metallicity (see Sect.~\ref{sec:chemicalmodel}) and were afterwards enriched by the surrounding donor stars. Because of the different star formation history of each ring, after $\sim 3.5$ Gyr of evolution, there are 808 and 443 stellar particles with   $Z > 0$  formed from target gas particles originally   belonging to the inner and outer rings, respectively. 

In Fig.~\ref{fig:maps}, we display both the initial gas density distribution (left panel) and the final stellar density distribution (right panel). The regions corresponding to the inner and outer rings are shown (purple and orange shaded regions, respectively), along with the positions of the target particles, displayed as empty circles for gas and filled circles for stars.

\begin{figure}[t]
\centering
\includegraphics[width=9cm]{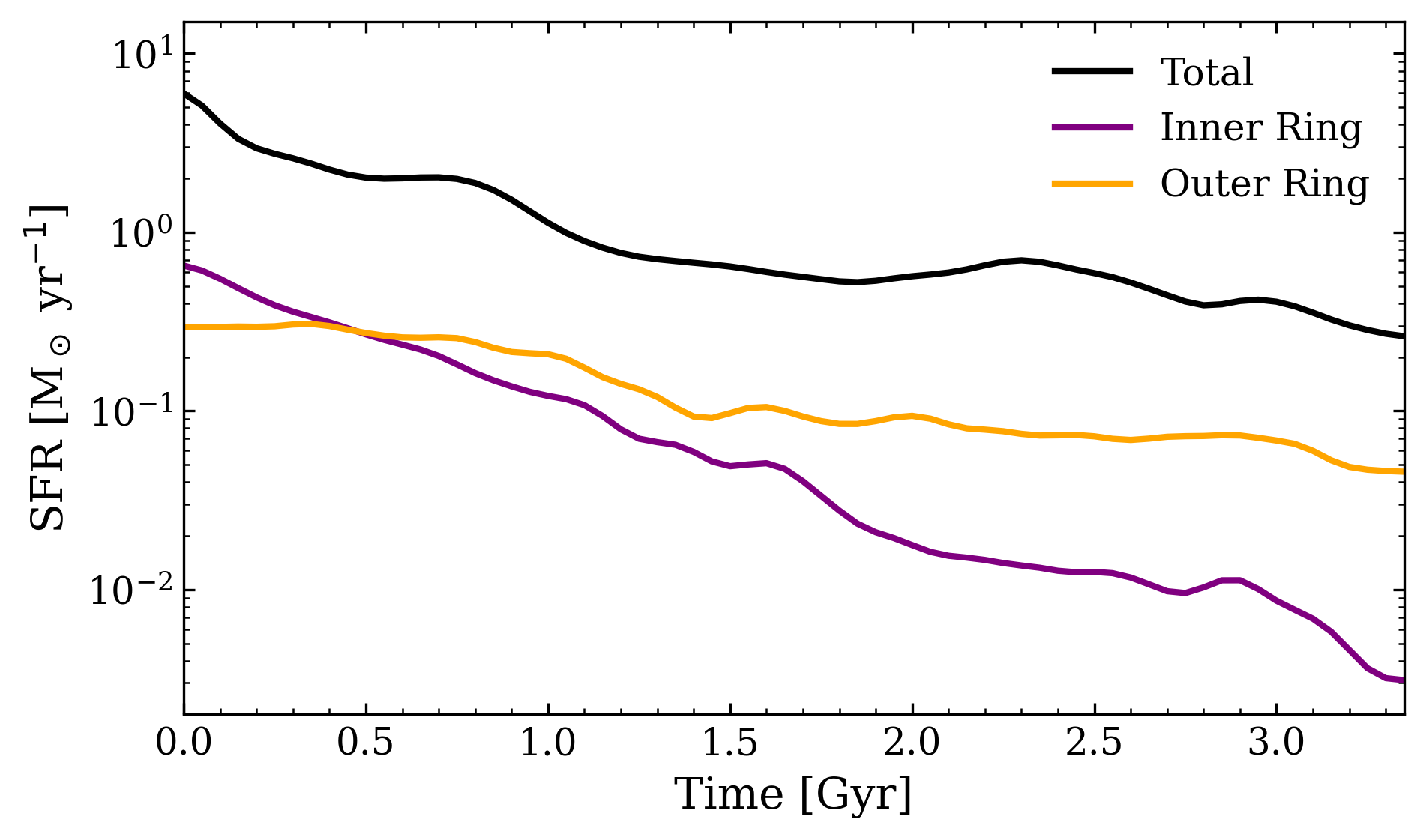}
\caption{Star formation-rate evolution for the whole galaxy (black line) and the inner and outer rings (purple and orange lines, respectively). After 3.5 Gyr of evolution, the star formation of the inner region decreases by almost three orders of magnitude while the outer region shows a more steadily decreasing star formation activity. 
\label{fig:SFH}}
\end{figure}

In Fig.~\ref{fig:SFH}, we present the star formation rate (SFR) of the entire galaxy, along with that measured within each of the two rings. Overall, the SFR declines over time, although a minor burst is evident around 1 Gyr. At early times, the inner ring exhibits a higher SFR than the outer ring, driven by the higher gas densities near the galactic centre. Rapid gas consumption, combined with bar-driven gas inflows toward the central regions, leads to a fast decline in star formation activity in the inner ring. In contrast, the outer ring exhibits a lower initial SFR, but it is more steadily sustained over time.

The spatial distribution of target particles at the final stage of the simulation (see right panel in Fig.~\ref{fig:maps}) reveals the impact of the bar, as most particles from the inner ring have migrated toward the central region. Only a few target particles of the inner ring moved outward. In contrast, most of the target particles from the outer ring remain there, with approximately 60\% still in the gas phase after 3.5 Gyr of evolution.

\section{Enrichment history of the inner and outer rings}
\label{sec:history}

\begin{figure}[t]
\includegraphics[width=0.5\textwidth]{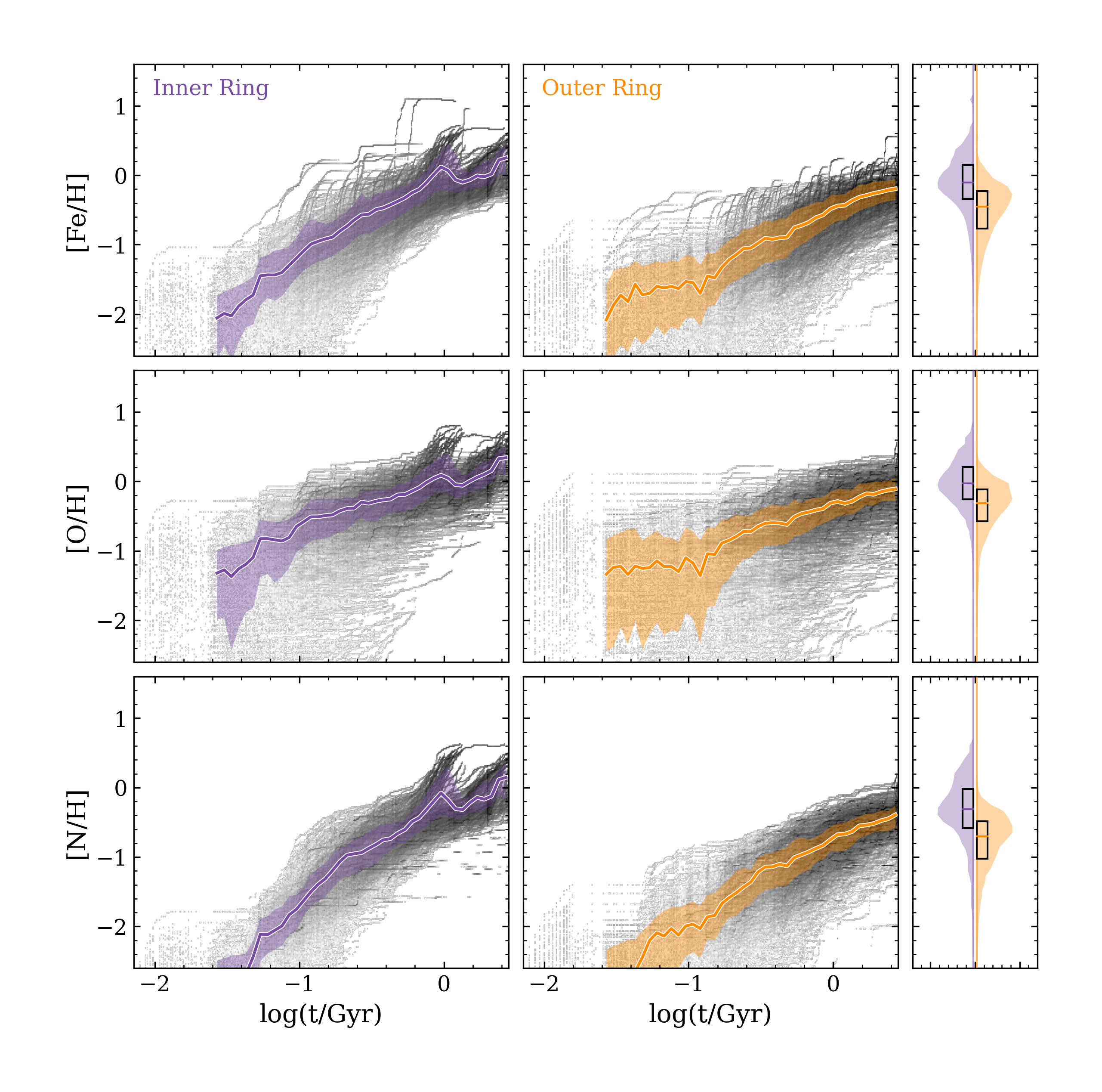}
\caption{Enrichment histories of target gas particles in the inner (purple, left column) and outer (orange, right column) rings. Three chemical abundances with different dominant stellar progenitors are shown: [Fe/H], [O/H], and [N/H] (top, middle, and bottom rows, respectively). For each ring, the background map illustrates the individual evolution of gas particles, while the solid line indicates the median abundance and the shaded regions correspond to the 25th and 75th percentiles. The right panels show violin plots of the [X/H] distributions with overlaid box plots indicating the median and interquartile range.
\label{fig:part_enrich_path}}
\end{figure}

Because of both the different star formation histories and the dynamical evolution, the target gas particles in the inner ring are more rapidly enriched to higher levels  than those in the outer ring.
This can be clearly seen in  Fig.~\ref{fig:part_enrich_path}, which displays the level of [O/H], [Fe/H], and [N/H]\footnote{We adopt the standard definition of [X/H] as the logarithmic ratio between the number densities of element $X$ and hydrogen, relative to their solar values, expressed in dex, i.e. [X/H] $= \log_{10}(n_X/n_H) - \log_{10}(n_X/n_H)_\odot$, where $n_X$ and $n_H$ denote the number densities of the corresponding elements.} abundances as a function of time for the target gas particles located in both rings until the time they become stellar particles.  We chose to show these three elements  since they have different main stellar progenitors for the adopted yields: SNII, SNIa, and AGB, respectively. 
We stress the fact that there are no diffusion mechanisms other than the mixing by injection and the dynamical evolution of the gas and stars. Although this approach may introduce some noise, it allows us to follow the origin of each chemical contribution produced by the three different stellar progenitors included in our chemical model.

From Fig.~\ref{fig:part_enrich_path}, we also find that the gas particles in the inner ring are enriched earlier than those in the outer ring. The central regions exhibit higher chemical enrichment, with [O/H] about 0.30 dex greater, and both [Fe/H] and [N/H] enriched by roughly 0.4 dex compared to the outer ring. From this figure, we can also see the delay in the injection of Fe and N with respect to O. The histories of enrichment are also modulated by the star formation history of the rings. The global trends are evident from the median values, but a diversity of behaviours is also observed at the particle level, as indicated by the scatter in the relation.

In order to quantify the rate of enrichment that each gas particle has before being converted into stars,  we define the chemical enrichment rate, denoted as $\mu_{\rm X}$ for a given chemical element X, as the total mass of that element received by a gas particle per unit time. This quantity captures the cumulative effect of all enrichment events experienced by a gas particle before its conversion into a stellar population. Given the variety of enrichment channels, namely, SNII, SNIa, and AGB stars, and their distinct timescales, the local star formation history, and the dynamical evolution of both gas and stars, $\mu_{\rm X}$ is expected to vary significantly between different elements and across different regions of the galaxy.
In Appendix~\ref{app:A}, we explore in more detail the correlations between the enrichment rates of different elements, as well as the intrinsic differences between the inner and outer rings. We will use  $\mu_{\rm X}$ to interpret our results in Section~\ref{sec:phylo}.

\begin{figure*}[ht!]
\centering
\includegraphics[width=\textwidth]{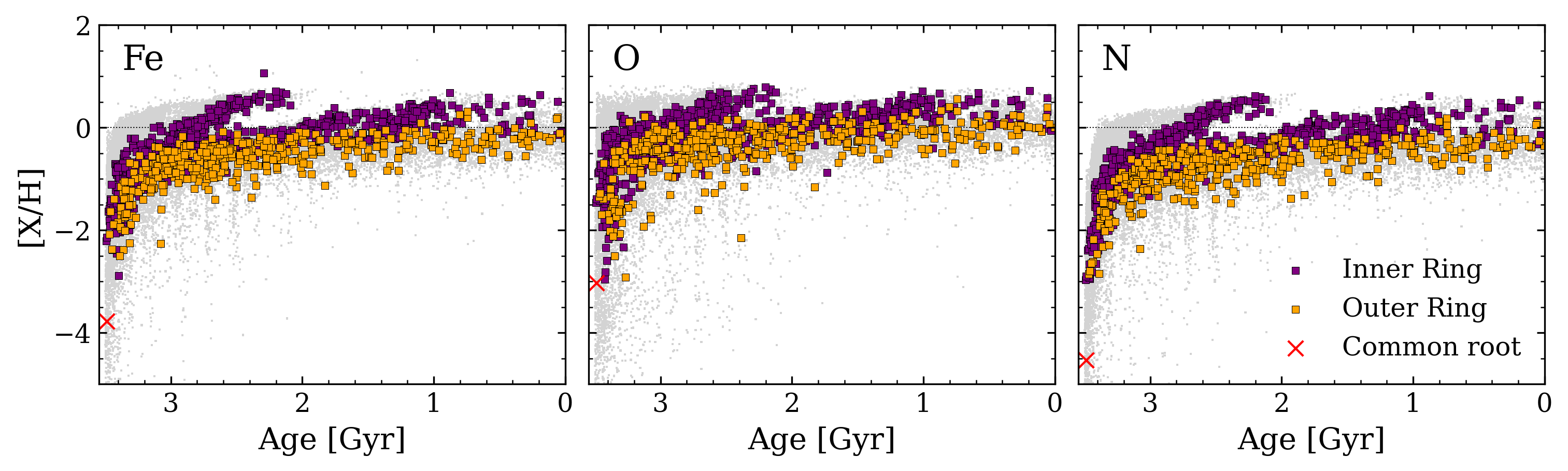}
\caption{Age-metallicity relation (AMR) of the stellar populations in the inner ring  (purple squares) and outer ring (orange squares). The global AMR is displayed in the background for reference (grey dots). Solar values are depicted by  horizontal dotted lines. The star particle taken as the root is also included (red cross).}
\label{fig:AMR}
\end{figure*}

Another way to examine the evolution of chemical abundances is through their variation as a function of stellar population age\footnote{The stellar age is obtained by subtracting the star formation time (SFT) from the final time up to which the simulation was run.}, as captured by the age–metallicity relation (AMR). The AMR serves as a fossil record for the cumulative enrichment of the ISM, as shown in Fig.~\ref{fig:part_enrich_path}.
In Fig.~\ref{fig:AMR}, we display the AMRs built using the three adopted chemical abundances. Stellar populations in the outer ring are younger than those in the inner ring, with a median age approximately half that of the inner population. As expected, the stellar populations in the inner ring exhibit systematically higher levels of chemical enrichment and a faster enrichment rate during the first few Myr compared to those in the outer ring. The AMR in the inner ring also displays a distinct double sequence, particularly evident in [Fe/H] and [N/H]. \textcolor{black}{The more metal-rich sequence arises from the injection of iron and nitrogen (through SNIa and AGB feedback) by older stellar populations that migrated from the inner ring into the very central region}\footnote{We define the central region of the galaxies as the region located inside the inner ring, i.e., at $r < 2$ kpc.}. These older stars formed in a central burst a few Gyr ago. Since the target particles in the outer ring do not migrate to the very central region and there are no pre-existing older populations, this effect is not observed and the AMR behaves as one smooth sequence.

Evidence for a bifurcated AMR in the MW has been found in both globular clusters \citep[see, e.g.][]{Leaman2013} and individual stars \citep[e.g.][]{Nissen2020}, mostly in the solar neighbourhood \citep{jofre2017}. Semi-analytical and fully analytical models, including one- and multizone chemical evolution frameworks, have shown that such bifurcations can naturally arise in two-infall scenarios, where the abundance is reset by accretion of metal-poor gas \citep{chiappini1997, Spitoni2021, Palicio2023}. In our isolated galaxy simulation, the emergence of two distinct AMR sequences is produced by the bar, which drives gas inflows from the inner ring into the central region at different epochs.

\begin{figure}[ht!]
\centering
\includegraphics[width=0.5\textwidth]{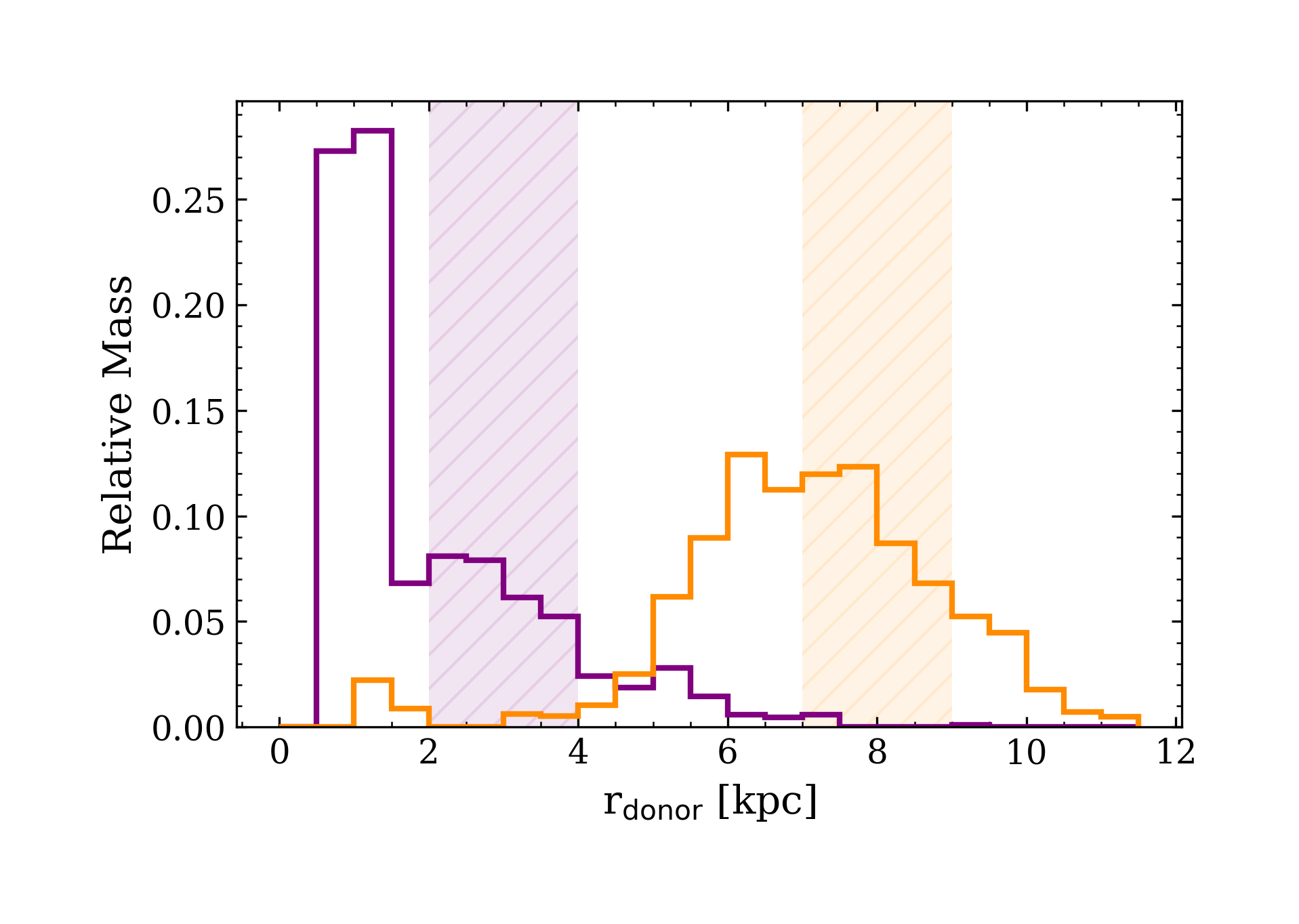}
\includegraphics[width=0.5\textwidth]{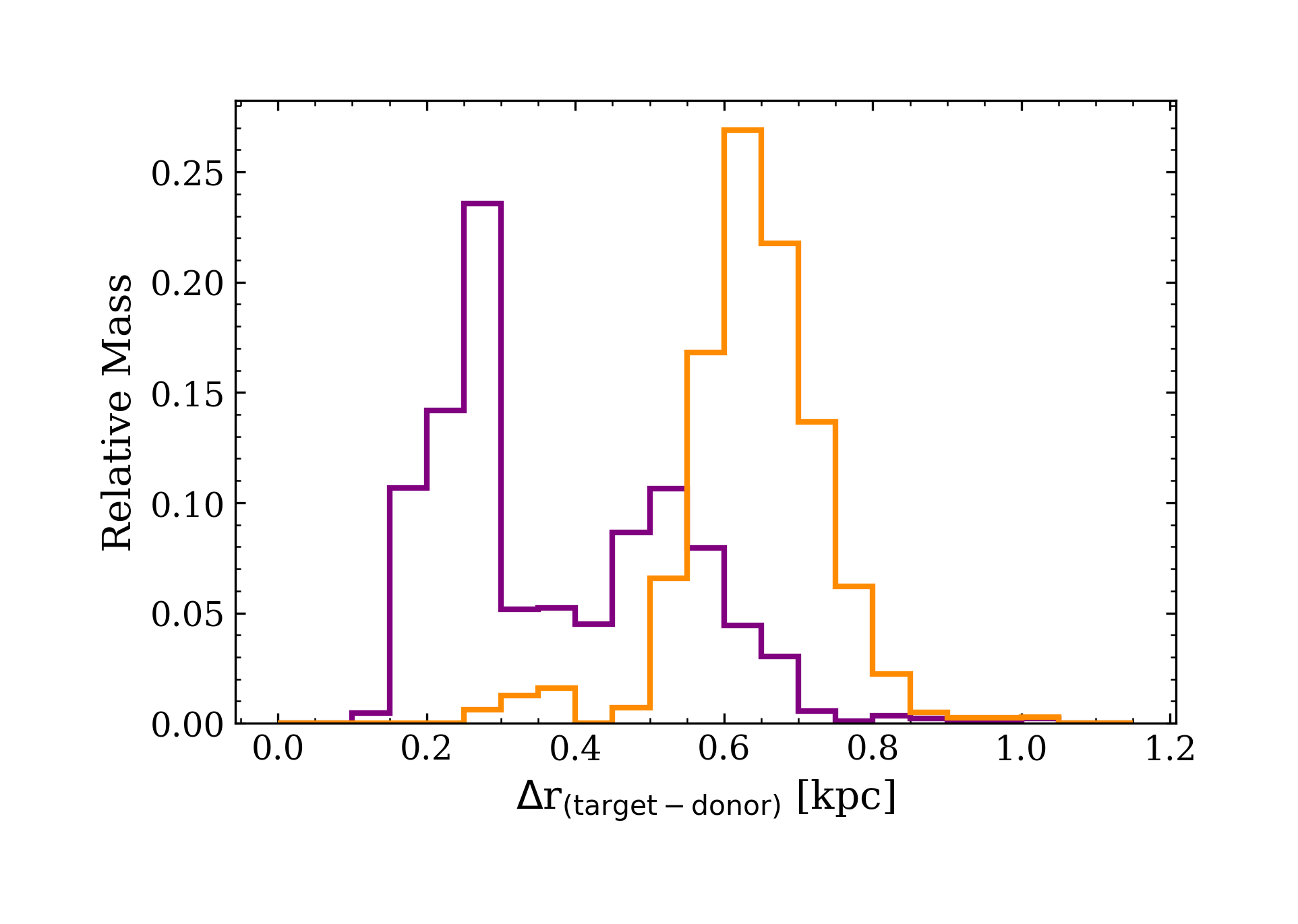}
\caption{ 
{
Donor-particle radial and relative-distance distributions. Upper panel: histogram of the radii of the donor particles for target gas particles in the inner (corresponding ring. The shaded regions correspond to the definition (in kpc) of each ring. Lower panel: distance between the target and the donor particles in the inner and outer rings.}
\label{fig:donors}}
\end{figure}

To characterise the enrichment process of the target SSPs in the selected rings, Fig.~\ref{fig:donors} shows the distribution of the stellar mass fractions contributed by individual donor particles. The upper panel presents these contributions as a function of the galactocentric radius of the donors, measured at the time they inject chemical elements into the target particles, while the lower panel shows the distribution as a function of the relative distance between each target SSP and its corresponding donors, also evaluated at the time of enrichment. In both panels, results are shown separately for the inner and outer rings. All distributions are normalised to the total stellar mass of the corresponding ring.

The bimodality in the distribution for the inner ring can be associated with the target particles that migrated into the central region (shorter distances) and those that either remained in the ring or, in a very small fraction of cases, migrated outwards. The outer ring shows no significant bimodality consistent with a more quiescent history of enrichment. The shaded areas in the upper panel denote the regions originally mapped by the inner and outer gas  samples. From these figures we can also appreciate that the gas particles in the outer ring, a region dominated by the spiral arms, receive chemical elements from wider than the original distribution. The larger distances between particles in the outer ring can also be explained because of the lower density in the disc than in the central region. We can visualise this from Fig.~\ref{fig:maps}.

The analysis presented in this section shows the cycle set by the injection of chemical elements from different stellar progenitors (donors) into the ISM, which is then converted into stars and retain the chemical pattern of the parent gas particle. This information provides a basis for interpreting the phylogenetic trees constructed under the hypothesis that chemical abundances behave as evolutionary traits, from which the history of chemical evolution can be reconstructed.

\section{Galactic phylogenetics}
\label{sec:phylo}
We constructed phylogenetic trees using the target SSPs from the inner and outer rings, whose chemical enrichment histories were discussed in the previous section.
A detailed discussion on the basis of galactic phylogenetics is given by \citet{jofre2017} and \citet{debrito2024}. 

Here, we briefly comment on the main concepts adapted to astrophysics. As mentioned in the Introduction, the structure of a phylogenetic tree is defined by its branching pattern, which reflects relationships among the stellar populations. In biological contexts, tips are the terminal nodes of the tree and correspond to the observed present-day species used to construct it. Internal nodes represent their most recent common ancestors. In our case, these tips correspond to present-day stellar populations characterised by their chemical abundances and ages. Although observed today, these populations retain the chemical properties of the ISM at their formation time. In this sense, all tips can be interpreted as fossil records of past ISM conditions.

Distance matrices provide a quantitative measure of dissimilarity between stars based on observed traits. In our case, these traits are the chemical abundance ratios of the stellar populations, expressed in the standard logarithmic notation [X/H] (see Sect.~\ref{sec:history}). Chemical elements are suitable phylogenetic traits because they are inherited and altered across successive generations through astration and ISM enrichment (`descent with modification'), which enables a direct connection to phylogenetic reasoning \citep{jofre2017}. We therefore construct a chemical distance matrix by computing the chemical distance\footnote{We use Euclidean distances in [X/H] space following previous works (see \citealt{debrito2024}).} between each particle and all others in the selected sample. Based on this matrix, we estimate phylogenetic trees using the neighbour-joining (NJ) algorithm \citep[e.g.][]{gascuel2006}. NJ is a distance-based method, hence, its branch lengths are estimated directly from the input distance matrix and thus inherit its units; consequently, branch lengths in our trees are expressed in dex units, and the chemical distance between any two leaves is obtained by summing the branch lengths along the path connecting them \citep[see the detailed discussion in][]{jofre2017}. 

Our trees are rooted. Rooting a tree is a critical component as the root represents the base of the phylogeny -- the earliest common ancestor from which all other lineages have evolved. Thus, by defining the root of the tree we can infer the direction of evolutionary relationships so that the more distant a branch is from the root the more evolutionary change separates it from the origin of the tree. Most biological rooting methods rely on specific evolutionary models that are not directly applicable to astrophysical contexts.
However, because we work with a simulated galaxy, we have knowledge of the formation times of all stellar particles. We therefore adopt a physically motivated approach by selecting \textcolor{black}{the oldest stellar particle, formed with a lower level of enrichment than the target particles}. The age and abundance of the selected stellar particle are shown in Fig.~\ref{fig:AMR} for reference. The root is shared by all the simulated trees throughout this work. We used the same root for both regions since they both start from primordial metallicity.

The root provides a meaningful temporal reference for the tree, under the assumption that existing stellar populations enriched the ISM, from which subsequent generations of stars formed, thereby establishing a hierarchical pattern of chemical evolution. In a closed-box system, it is assumed that the ISM is continuously enriched by the existing stellar populations, leading to newly formed stars with increasingly higher metallicities. This process naturally establishes a hierarchy, from old, metal-poor populations to young, metal-rich ones. This hierarchy can be broken when material is allowed to flow in or out of the system. Hence, this assumption is rarely valid in a cosmological context of galaxy formation, where processes such as gas infall, outflows, and stellar redistribution (e.g. migration) can break the chemical hierarchy. These mechanisms introduce complex features in the NJ trees that require more detailed analysis, best addressed through numerical simulations.

Regarding the number of stellar particles used to build the phylogenetic trees, as discussed by \citet{debrito2024}, although there are sample-to-sample variations, a minimum of 200 stellar particles is sufficient to construct trees that significantly deviate from a random distribution of chemical abundances. In this work, we adopt 400 stellar particles per region, randomly selected from the full sample of target particles. In the following section, we present a representative example of such a tree for each ring. The size of the underlying sample ensures that the inferred trends are representative of the full stellar population in each region, while remaining sufficiently limited to prevent saturation of the phylogenetic trees.

With these trees, we aim to explore the information they encode. In particular, the complex evolution of the central region, including the formation of the bar, offers the opportunity to investigate how this dynamical evolution is imprinted on the chemical abundances of the stellar populations and reflected in the structure of the phylogenetic trees. 

\subsection{Phylogenetic trees of the target SSPs in the inner ring}

The gas particles selected as targets in the inner region followed different evolutionary paths as already discussed. The central region of the disc developed a bar due to internal instability, as can be seen from Fig.~\ref{fig:maps}.  As expected, the bar triggers gas inflows and, as a result, 64\% of the inner ring mass fraction fell along the bar.

\begin{figure*}[ht!]
\centering
\includegraphics[width=0.91\textwidth]{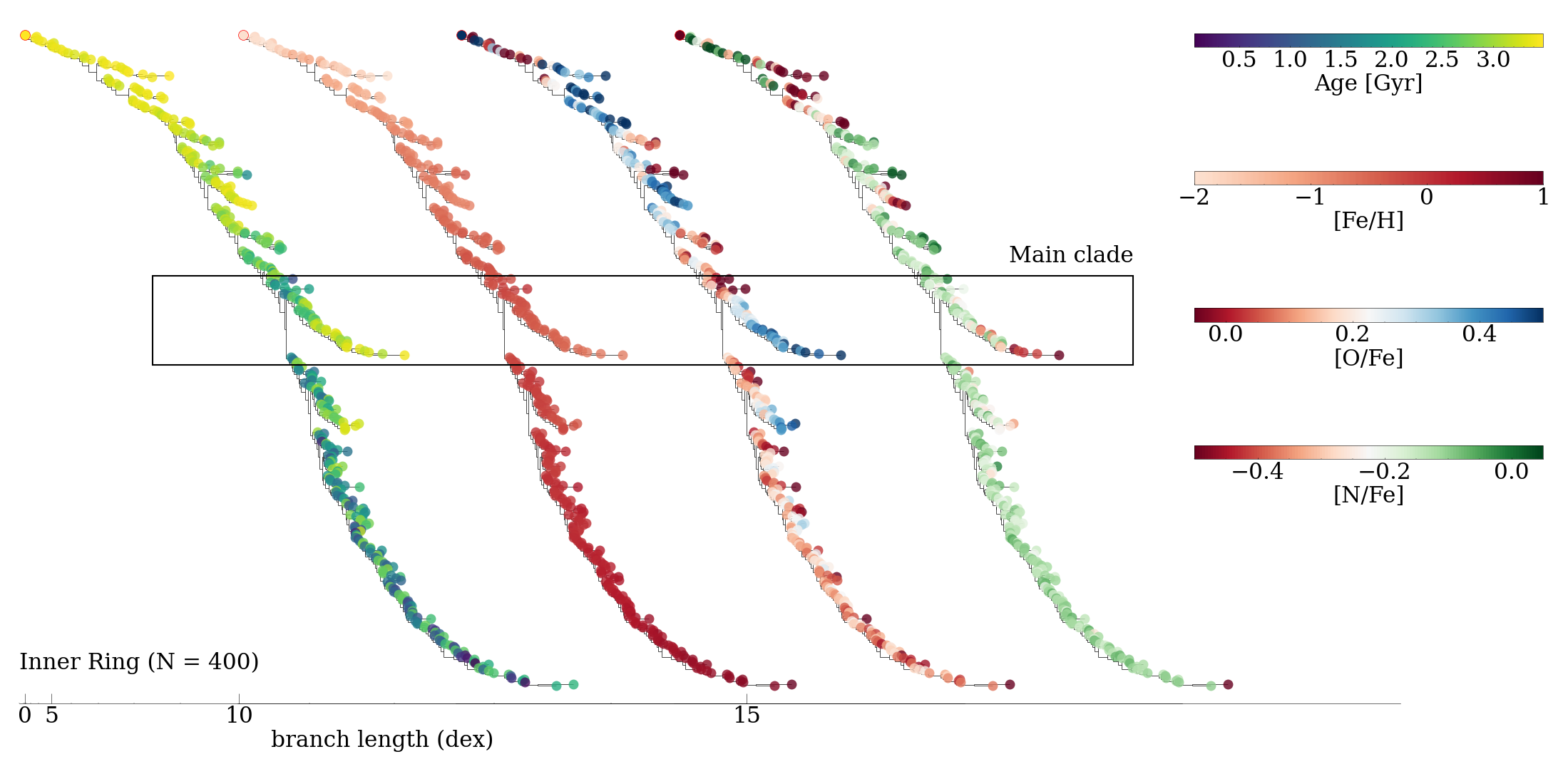}
\caption{ 
\label{fig:innertree}
Phylogenetic tree constructed with 400 target stellar particles randomly selected from the inner ring. Colour indicates the age of the stellar particles (left), their iron abundance ([Fe/H]; centre), and their oxygen and nitrogen abundance ([O/Fe] and [N/Fe]; right). Branch lengths are shown to scale only in the leftmost tree; in the other panels the trees are laterally shifted for easier visual comparison. The x-axis is displayed on a power-law scale to enhance visual clarity. The boxed region highlights the main clade discussed in the text. The root is located at the top left of each tree and is highlighted by a red contour.}
\end{figure*}

The phylogenetic tree of the inner region is displayed in Fig.~\ref{fig:innertree}. Each panel shows the same tree coloured by stellar age, [Fe/H], [O/Fe], and [N/Fe]. Globally, we observe that younger SSPs tend to exhibit systematically higher [Fe/H], lower [O/Fe], and higher [N/Fe], \textcolor{black}{as expected from the standard chemical enrichment arguments discussed above}. However, the tree does not present a strict caterpillar structure particularly evident in [Fe/H], displaying a systematic variation of age and metallicity from the root to the most distant SSPs \citep[see][for a more extended discussion]{debrito2024, jofre2025}. Instead, it features noticeable clustering of SSPs, which we refer to as branches or clades. In evolutionary biology, clades often indicate a distinct population of species sharing a common ancestry. In the astrophysical context, these clades suggest the presence of groups of stars with a common enrichment history that deviates from the main hierarchical trend. Notably, within these clades the expected age gradient is disrupted, such that some branches include SSPs located further from the root that are, in fact, older.

In particular, one prominent clade (dubbed \textit{main clade}, see Fig.~\ref{fig:innertree}) is composed predominantly of old SSPs. This apparent inversion in the age hierarchy arises because the tree is constructed solely based on chemical abundances, without incorporating temporal information. These stars exhibit intermediate [Fe/H], high [O/Fe], and low [N/Fe], consistent with being formed from an ISM enriched predominantly  by SNII linked to current star formation activity, but which was already polluted by SNIa and AGB stars. \textcolor{black}{This chemical pattern suggests that these SSPs formed during an early, bursty star formation episode, consistent with a rapid injection of iron and oxygen.} We also note that smaller clades exhibit age and abundance-ratio patterns similar to those of the main clade, suggesting analogous processes operating on smaller scales or being under-represented in the random sample.

The recovery of the hierarchical structure is consistent with the action of gas inflows and material redistribution driven by the bar. Farther from the root, at higher metallicities, enrichment is dominated by older stars in the central region, whose delayed feedback enriches the ISM primarily through SNIa and AGB channels. As a result, these SSPs show lower [O/Fe] and higher [N/Fe], and the tree presents a strong hierarchical pattern. 

\textcolor{black}{Younger, more chemically enriched stars are distributed in a more hierarchical manner for the three selected chemical elements. However, a mix of stellar ages is found, resulting from the efficient mixing of chemical elements and the influence of bar-driven inflows.}

\subsection{Phylogenetic trees of the target SSPs in the outer ring}
\begin{figure*}[ht!]
\centering
\includegraphics[width=0.91\textwidth]{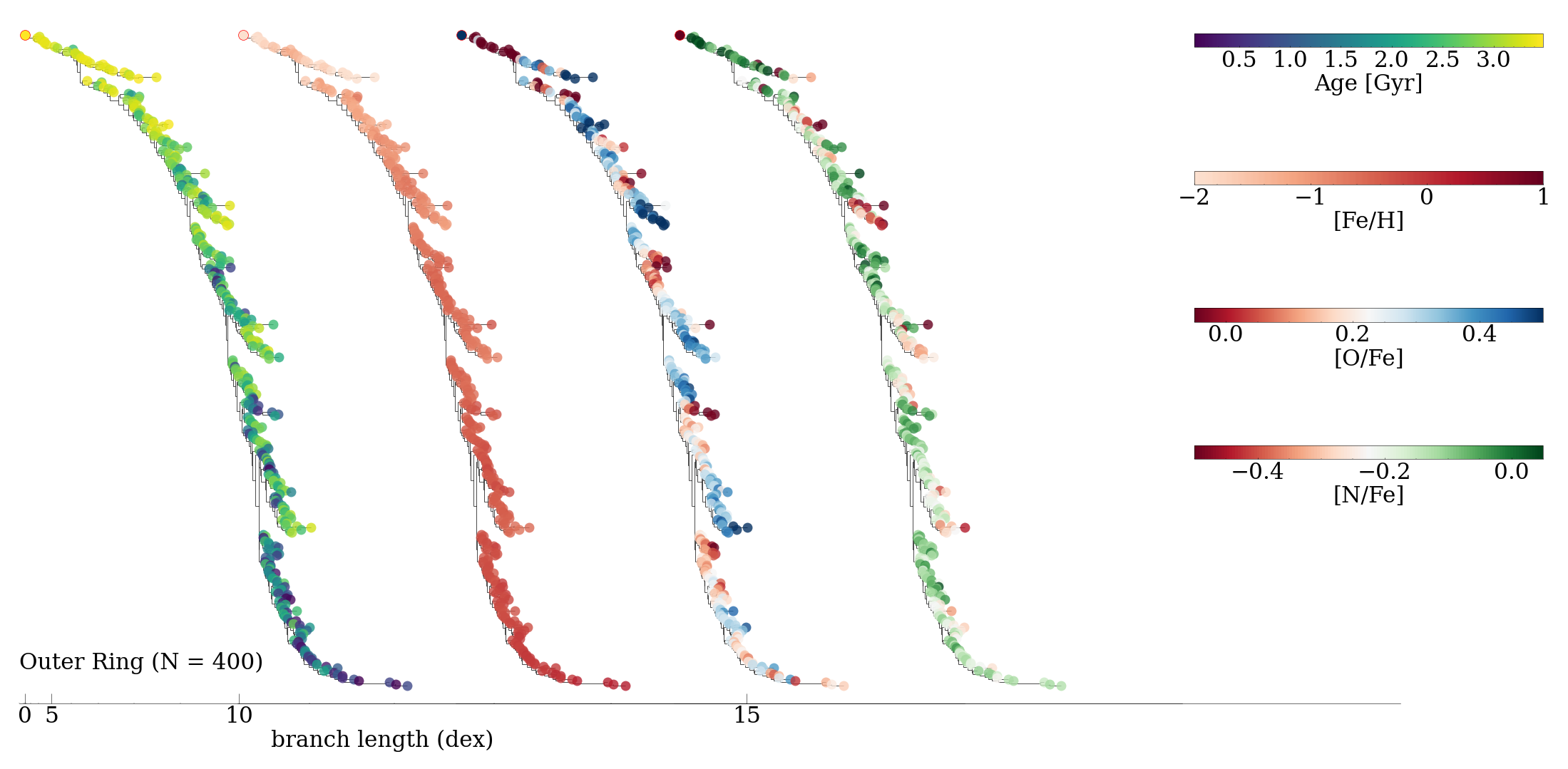}
\caption{ 
\label{fig:outertree}
Phylogenetic tree constructed with 400 stellar particles randomly selected from the outer ring. The root is common to the inner- and outer-ring trees and is located at the top left of each tree and highlighted by a red contour.}
\end{figure*}

Figure~\ref{fig:outertree} shows the phylogenetic tree for SSPs in the outer ring. Here, the tree exhibits a more caterpillar-like structure. Nevertheless, some short clades are present, indicating mild departures from symmetry, likely associated with the formation of spiral arms and substructures along them. The ISM in this region appears more inhomogeneous, as reflected by the presence of these shorter clades. This inhomogeneity arises in part from the different production and ejection timescales of oxygen, iron, and nitrogen. These enrichment sources operate on distinct timescales that are strongly dependent on both stellar mass and metallicity. Since our simulations account for these effects in a self-consistent manner, differential mixing is naturally expected.

In addition, gas and stars follow their own dynamical evolution. The formation of the bar and spiral arms induces large-scale motions that mix both components, shaping the spatial distribution of the injected elements. Star formation occurs preferentially along spiral arms and in clumps, introducing further complexity to the chemical enrichment patterns. These dynamical features modulate the efficiency and spatial scale of chemical mixing, thereby contributing to the observed chemical inhomogeneities.

We find that, depending on the strength of star formation and the importance of central gas inflows, distinct imprints can emerge in the phylogenetic trees. In the case of the inner ring, this is reflected both in the presence of a well-populated clade and in the strong hierarchical structure observed near the end of the tree. In contrast, regions that follow a more steady evolutionary path—characterised by sustained star formation and gradual mixing—display phylogenetic trees that encode a smoother enrichment history. These trees exhibit an overall hierarchical pattern of chemical evolution, populated by short clades that reflect a more chemically heterogeneous ISM.

\begin{figure*}[ht!]
\centering
\includegraphics[width=0.91\textwidth]{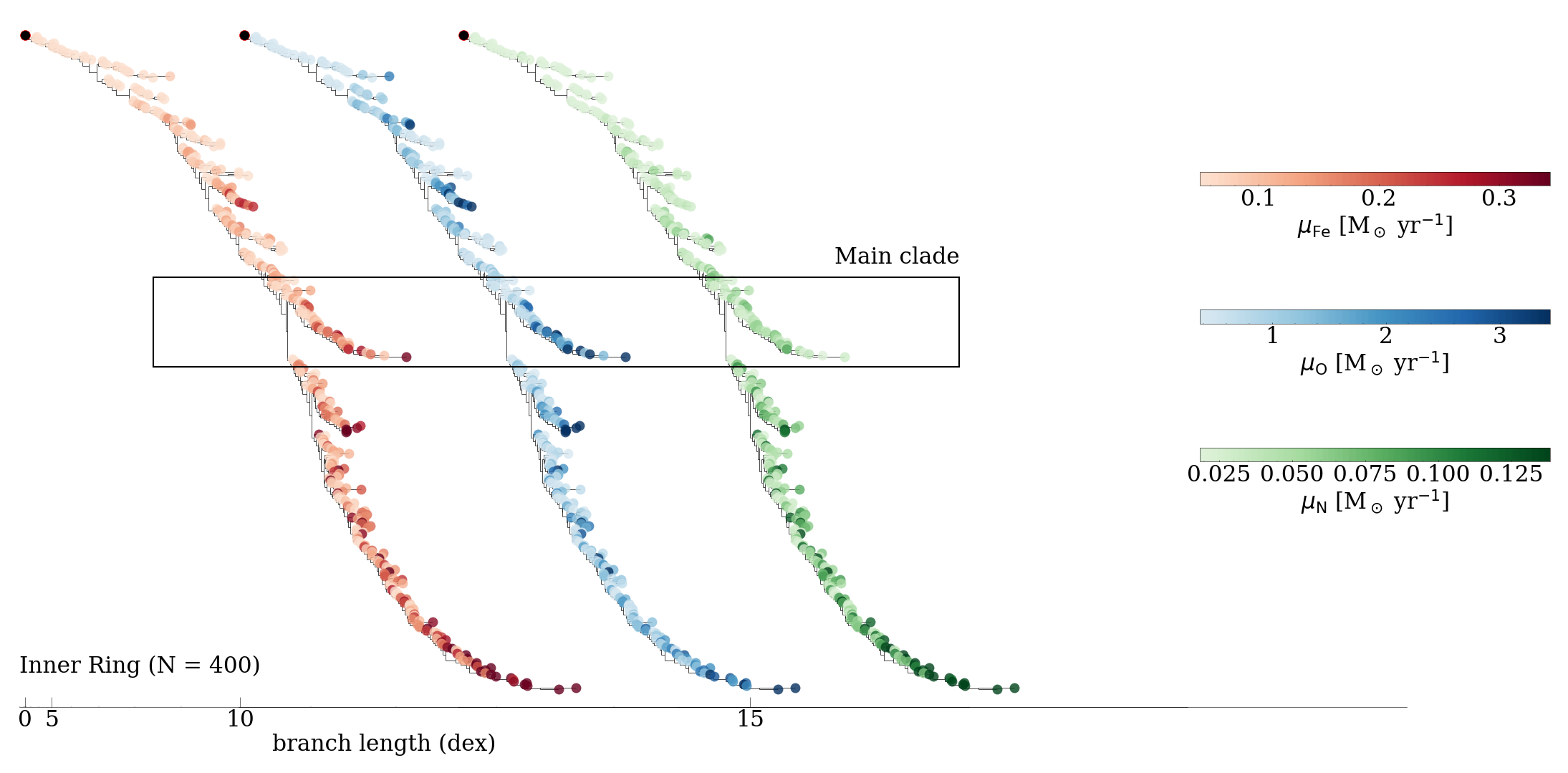}
\caption{ 
\label{fig:innertreeslopes}
Phylogenetic tree of the inner ring (same as in Fig.~\ref{fig:innertree}) coloured by the chemical enrichment rate of iron (left tree), oxygen (central tree), and nitrogen (right tree). The boxed region highlights the main clade discussed in the text.}
\end{figure*}

\begin{figure*}[ht!]
\centering
\includegraphics[width=0.91\textwidth]{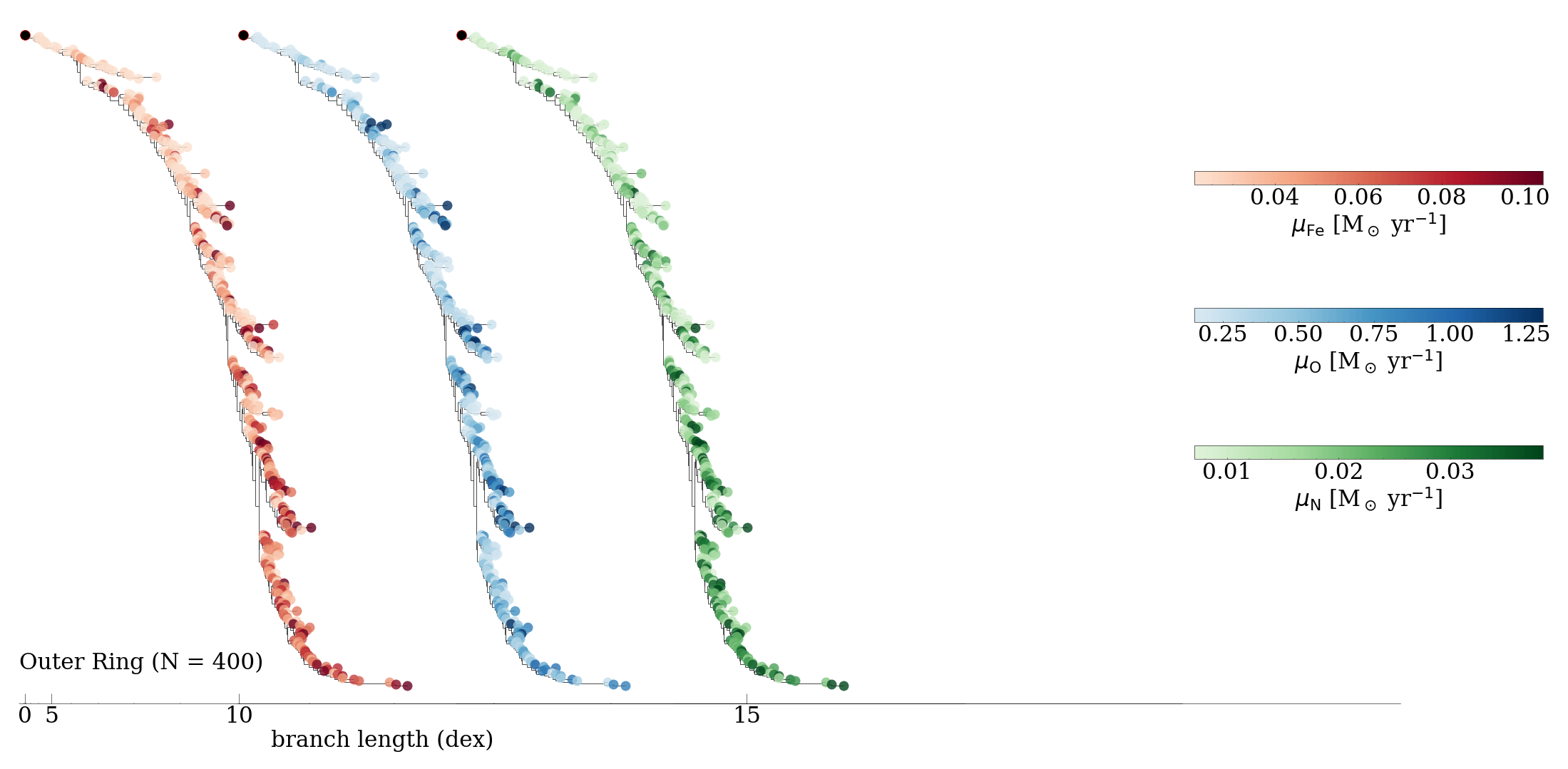}
\caption{ 
\label{fig:outertreeslopes}
Phylogenetic tree of the outer ring (same as in Fig.~\ref{fig:outertree}) coloured by the chemical enrichment rate of iron (left tree), oxygen (central tree), and nitrogen (right tree).}
\end{figure*}

\begin{figure}[ht!]
\centering
\includegraphics[width=0.46\textwidth]{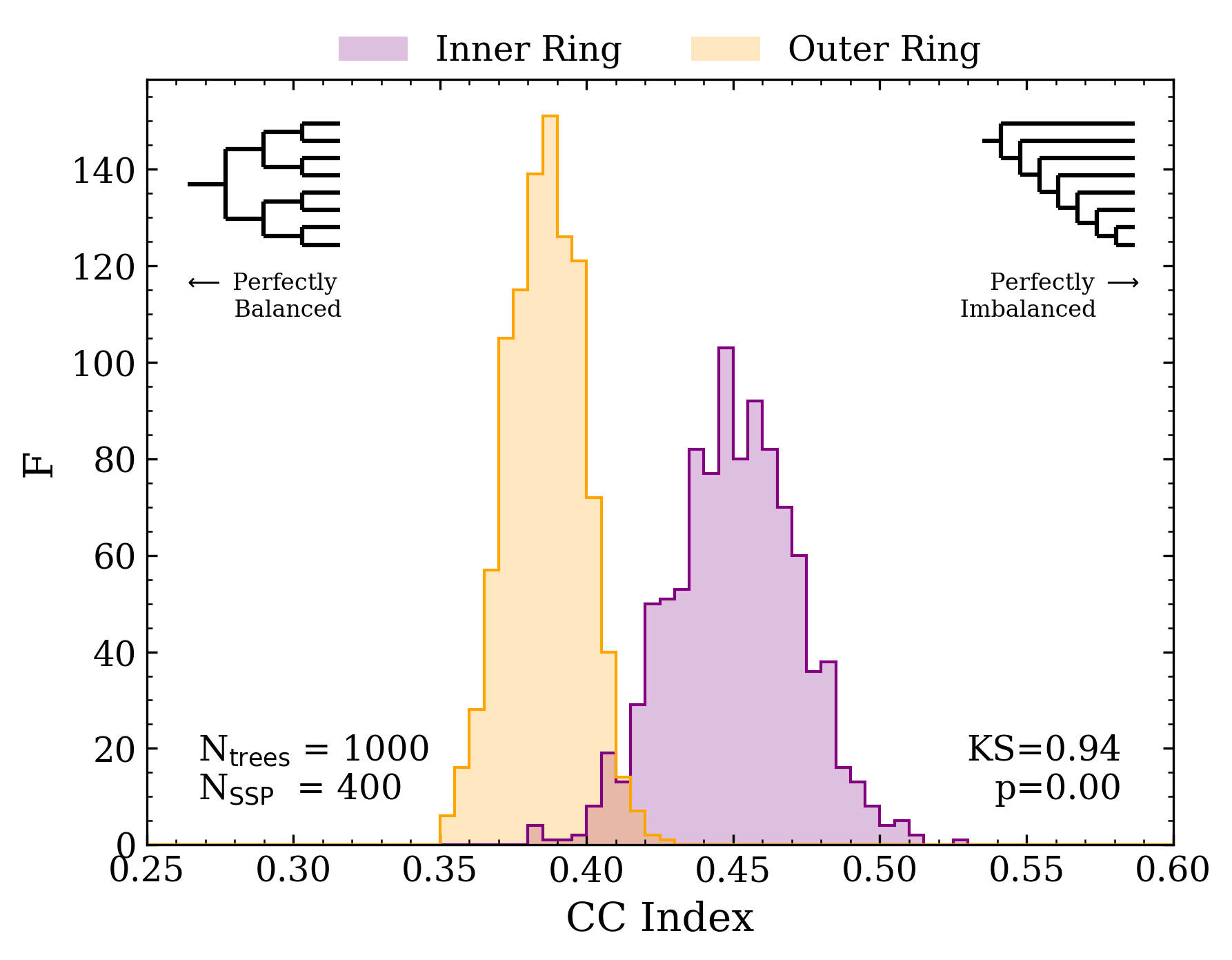}
\caption{
    \label{fig:indices}
    Distribution of the tree balance metric, CC index, computed for a sample of 1000 phylogenetic trees built from 400 SSPs randomly selected from the inner (purple) and outer (orange) rings.}
\end{figure}

\begin{figure}[ht!]
\centering
\includegraphics[width=0.46\textwidth]{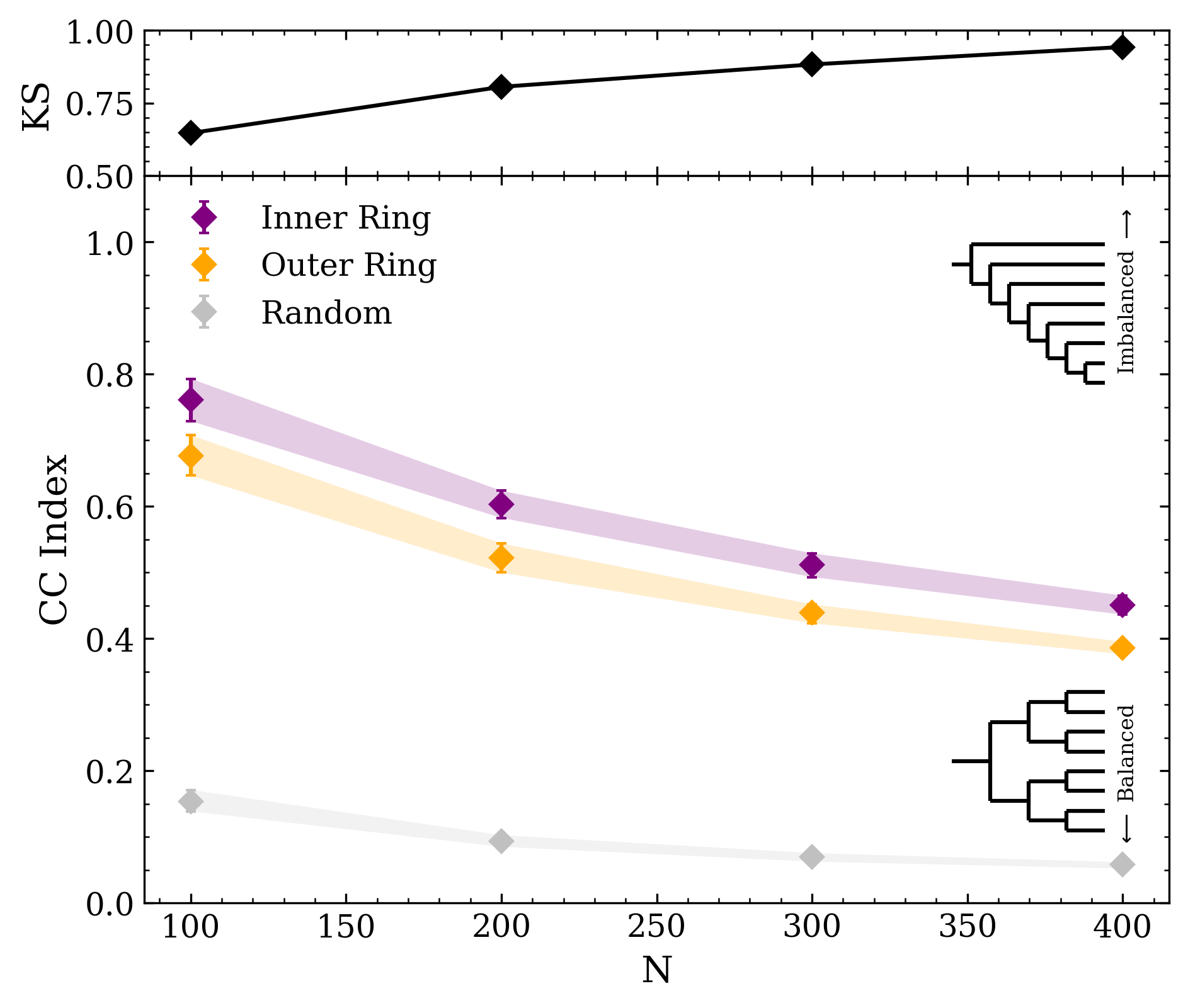}
\caption{
    Convergence of the CC tree-shape index as a function of the number of stellar particles adopted to build the tree (N$_{\rm SSP}$). Shaded regions depict the 25th–75th percentile range. The upper sub-panel displays the Kolmogorov–Smirnov (KS) statistic between the inner and outer ring distributions at each N$_{\rm SSP}$; $p$-values are close to zero in every case.
}
\label{fig:convergence}
\end{figure}

\subsection{Phylogenetic trees of mixed populations}

To assess the robustness of the inferred phylogenetic signal, we constructed phylogenetic trees from mixed samples combining stellar populations from the inner and outer rings. The details can be found in Appendix~\ref{app:C}.  We find that the resulting tree structure differs significantly from those obtained for the individual regions, reflecting the intrinsic chemical contrasts between the two parent populations. In particular, clades identified in the individual trees are not always preserved in the mixed case, as chemically similar populations from different regions can redistribute along the tree, leading to the emergence of new dominant clades and the partial dissolution of pre-existing ones. Statistical analysis based on the Colless imbalance index (see Sect.~\ref{subsec:comparison}) shows that the mixed trees exhibit an intermediate degree of imbalance between the inner- and outer-ring cases, while remaining statistically distinct from both. These results highlight the sensitivity of phylogenetic reconstructions to sample composition. However, the phylogenetic signal remains consistent with the evolutionary history of the selected populations; the main challenge lies in disentangling this signal adequately. We will further explore this aspect in a forthcoming paper. In Appendix~\ref{app:C}, we also highlight an important limitation of the NJ method: proximity in the tree can also reflect chemical similarity and shared enrichment pathways, rather than exclusively direct ancestor--descendant sequences.

The analogy with biological phylogenies must therefore be treated with caution. In biology, the relationship between tips and internal nodes can often be interpreted genealogically; however, it is not always true (see Appendix~\ref{app:C} for more details).  In galactic applications based on chemical distances, the tree is better understood as a structured representation of chemical enrichment histories shaped by star formation, delayed feedback, mixing, and migration. Accordingly, branches and clade-like groups are especially informative about chemically distinct enrichment pathways, but they should not always be read as strictly time-ordered evolutionary endpoints.

\section{Discussion}
\label{sec:discussion}
\subsection{Rate of chemical enrichment}

Here we make use of the chemical enrichment rates, $\mu_{\rm X}$, defined in Section~\ref{sec:history}. These rates encode valuable information about the timing, origin, and efficiency of enrichment, particularly in regions with complex dynamical histories. For instance, an early, rapid enrichment dominated by SNII should manifest as a sharp increase in $\mu_{\rm O}$, while a slower, delayed contribution from SNIa and AGB stars would shape $\mu_{\rm Fe}$ and $\mu_{\rm N}$ differently. These differences encode aspects of the enrichment history not directly visible in the present-day abundances (see also Appendix~\ref{app:A}). 
Figures~\ref{fig:innertreeslopes} and~\ref{fig:outertreeslopes} show the same phylogenetic trees presented in Figs. \ref{fig:innertree} and~\ref{fig:outertree}, but  colour-coded by the enrichment rate $\mu_{\rm X}$ of the three representative elements: iron, oxygen, and nitrogen. These enrichment rates provide a complementary view of the chemical histories of the stellar populations.

In the inner ring (Fig.~\ref{fig:innertreeslopes}), the main clade shows relatively low enrichment rates in nitrogen, but enhanced $\mu_{\rm O}$ values followed by $\mu_{\rm Fe}$, consistent with rapid early enrichment by SNII-dominated feedback. This supports our previous interpretation that these stars formed during an early starburst phase, followed by prompt oxygen injection and limited contribution from delayed sources. \textcolor{black}{Towards the end of the tree}, where younger populations are found, the enrichment rates for nitrogen and iron are significantly higher, reflecting the growing contribution from SNIa and AGB stars as the system evolves embedded in the central region.

In contrast, the outer ring tree (Fig.~\ref{fig:outertreeslopes}) displays a smoother distribution of enrichment rates along the tree. While short clades exist, their enrichment patterns are more continuous, with fewer sharp contrasts. This is consistent with the more gradual and extended star formation history inferred from the [Fe/H], [O/Fe], and age distributions in Fig.~\ref{fig:outertree}. The presence of overlapping enrichment rates at similar branch length again reflects the stochastic nature of enrichment in a more chemically heterogeneous ISM. 
This is particularly evident in the short clades, where the enrichment rates are similar for the three tracers in contrast to the inner ring. \textcolor{black}{This shows the coexistence of different stellar populations with different ages, which is explained by the fact that the outer ring shows a more continuous star formation history (Fig.~\ref{fig:SFH}).}

 Taken together, the enrichment rates reinforce the view that the structure of the phylogenetic trees is shaped not only by the absolute abundance values, but also by the underlying temporal sequence and strength of chemical contributions from different channels (see also Appendix~\ref{app:A}).

\subsection{Comparing phylogenetic trees: tree balance}
\label{subsec:comparison}
To enable a more robust comparison of the trees from the two regions, we aim to adopt approaches commonly used in biology to quantitatively evaluate phylogenetic trees. In particular, we compute the Corrected Colless (CC) index \citep[][see also \citet{fischer2023tree}]{Heard1992, Kirxpatrick1993}, a standard measure of tree imbalance. This index quantifies the asymmetry of a rooted tree, with higher values indicating more imbalance. In our context, a higher CC index can be interpreted as a more hierarchical chemical enrichment history, where certain evolutionary paths dominate the formation of stellar populations. The measure of tree balance offers a complementary insight into the structure and assembly history encoded in the phylogenetic trees, reflecting the asymmetry and hierarchical nature of enrichment.

In order to compute this index, we generated 1000 random phylogenetic trees for the inner and outer SSP samples. Fig.~\ref{fig:indices} shows the distribution of the CC index for trees constructed from inner (purple) and outer (orange) ring particles. As expected, the trees from the inner ring tend to be more imbalanced, reflecting the strong hierarchy observed within the main clade, and at the metal-rich end of the tree (see Fig.~\ref{fig:innertree}). This supports the interpretation that the inner region experienced a more structured and burst-driven enrichment history.

We performed a Kolmogorov–Smirnov (KS) test to compare the distributions of the CC index between the inner and outer ring trees. The KS test yields $p$-values $\sim 0$, confirming that the distributions are statistically distinct. This demonstrates that the differences in tree morphology between the two regions are robust, even when using randomly selected subsets of stellar populations.

It is important to note that other tree–comparison techniques commonly used in biology \citep[see, e.g.,][]{fischer2023tree} are not straightforward to apply in astrophysics, given the different nature of the data and the assumptions on evolutionary traits. This underscores the need to adapt some of these metrics to the astrophysical context, or alternatively to develop new, more general methodologies based on the intrinsic properties of the trees themselves (Signor et al., in prep.). 

\subsection{Tree balance convergence with N$_{\rm SSP}$}

We also addressed the robustness of the tree balance metric—CC index—as a function of the number of stellar particles used to build the phylogenetic trees (N$_{\rm SSP}$). This analysis is essential to assess how sensitive the inferred evolutionary features are to sampling size, and whether the morphological differences between the inner and outer ring trees persist across different realisations.

Figure~\ref{fig:convergence} shows the variation of the CC index as a function of N$_{\rm SSP}$, for values ranging from 100 to 400. The shaded regions indicate the 25 and 75 percentiles across the 1000 random realisations per region. For comparison, we also include a control sample generated from randomly generated SSPs\footnote{The random sample was created by assigning, for each chemical element, a random value drawn from a uniform distribution bounded by the minimum and maximum values measured across the entire target sample.} (grey points), which serves as a baseline representing chemically unstructured populations.

\textcolor{black}{We observe that the CC index decreases with increasing N$_{\rm SSP}$ for all cases. This suggests that larger samples tend to produce more balanced trees. The effect likely arises because clades become more populated, which reduces the apparent asymmetry introduced by sampling noise when only a small number of particles is used.} However, across all N, the inner ring trees consistently show higher CC values than the outer ring, confirming that the inner region retains a more imbalanced structure regardless of sample size. In contrast, the random sample remains nearly flat and close to zero, as expected from trees with no internal chemical structure, expected to be perfectly balanced.

The upper small panels show the outcome of the Kolmogorov–Smirnov (KS) test performed between the inner and outer ring distributions at each N. The consistently high KS values ($> 0.75$; with $p$-values $\sim 0$ in every case) demonstrate that the differences between the two populations are statistically significant even for small samples (N$_{\rm SSP} = 100$), and remain robust as N$_{\rm SSP}$ increases. This confirms that the morphological signatures identified in the trees are not sensitive to N$_{\rm SSP}$ adopted for the sampling as long as N$_{\rm SSP} >100$ \citep{debrito2024}. Hence the chemical enrichment histories of the inner and outer rings stored by the phylogenetic trees are genuinely distinct between each other and different from a random metallicity distribution.

\section{Conclusions}

We have applied the phylogenetic method to a galaxy evolution simulation, to assess how it represents the evolution of stellar populations.
Our simulation followed the chemical evolution of an isolated disc and we tracked the enrichment of a sample of target gas particles. Our simulation followed the chemical evolution of an isolated disc, and we tracked the enrichment of a sample of target gas particles. Once converted into stellar particles according to the subgrid physics, they stored the chemical patterns acquired by their parent gas particles.
Two regions were tracked, one close to the centre and the other one in the disc, where spiral arms developed. 
A bar structure appeared during the first stages of evolution, which had an impact on the dynamics of the surrounding ISM producing gas inflows. This gas followed a different history in comparison to the gas that remained in the original selected region.
The outer ring gathers stellar populations that were affected by the formation of arms and a more local exchange of chemical elements with a weak azimuthal dependence. 
We analysed the phylogenetic trees of each region and found the following results.

\begin{itemize}

\item In the inner ring, the phylogenetic tree exhibits a well-defined clade composed of old stars enriched primarily by SNII. This structure reflects a strong star formation episode as probed by the enhanced oxygen abundances and high enrichment rates. At higher metallicities, the tree shows increasing contributions from SNIa and AGB stars, resulting in a hierarchical pattern.

\item The outer ring shows a more continuous, caterpillar-like, and symmetric tree structure, with smaller clades and a smoother gradient in enrichment rates and abundances. This is consistent with a more extended and steady star formation history, supported by local mixing processes associated with spiral arms.

\item The chemical enrichment rates ($\mu_{\rm X}$) provide an additional layer of interpretation for the tree structures. In the inner ring, they confirm a rapid early star formation episode (i.e. quick chemical enrichment) that gave rise to a distinct clade, while the outer ring displays smoother and more uniform rates, consistent with gradual and spatially extended enrichment.

\item We quantified tree morphology using a standard metric for tree balance: the Corrected Colless (CC) index. Trees from the inner ring show higher CC values, consistent with an early, hierarchical chemical evolution. Outer ring trees are more symmetric, reflecting recent, gradual enrichment. This difference is statistically robust across 1000 random realisations.

\item We also assessed the convergence of this index as a function of the number of stellar particles used to build the trees. The CC index shows stable trends, and the differences between inner and outer regions remain significant even for small samples (N$_{\rm SSP}$ = 100), as confirmed by high KS statistics.

\end{itemize}

These results show that galactic phylogenetics is a promising  and complementary tool for interpreting the fossil record of stellar populations, capable of reconstructing main features of the evolutionary pathways of distinct galactic regions purely from chemical information. This is important considering the difficulties involved in a robust estimation of stellar ages.  We acknowledge the fact that the NJ algorithm orders stellar particles by similarities in their chemical patterns and does not assume any prior information. More sophisticated methods based on networks could be powerful to study complex systems like galaxies. However, from a numerical perspective, our findings pave the way for interpreting the tree-based analyses emerging from stellar surveys, proving a more efficient tool that does not require further model fitting.

\begin{acknowledgements}
       We sincerely thank the referee for their insightful comments and suggestions, which significantly improved this manuscript. BTC gratefully acknowledges funding by ANID (Beca Doctorado Nacional, Folio 21232155). PBT acknowledges partial funding by FONDECYT-ANID 1240465/2024 and Núcleo Milenio ERIS NCN2021\_017. ES acknowledges funding by FONDECYT-ANID Postdoctoral 2024 Project N°3240644 and thanks the Núcleo Milenio ERIS NCN2021\_017. PJ is supported by FONDECYT Regular 1231057. TS acknowledges financial support from Inria Chile ANID project CTI230007. PD is supported by a UKRI Future Leaders Fellowship (grant reference MR/S032223/1). CAG acknowledges support from FONDECYT Iniciación 11230741. This project has received funding from the European Union Horizon 2020 Research and Innovation Programme under Marie Skłodowska-Curie Actions (MSCA) grant agreement No. 101086388-LACEGAL. We acknowledge partial support by ANID BASAL project FB210003. This project used the Ladgerda Cluster (FONDECYT 1200703/2020 hosted at the Institute for Astrophysics, Chile), the NLHPC (Centro de Modelamiento Matem\'atico, Chile), and  Geryon clusters (Center for Astrophysics, CATA, Chile). 
\end{acknowledgements}

\bibliographystyle{aa} 
\bibliography{biblio}  

\begin{appendix}
\onecolumn

\section{Rates of chemical enrichment and abundance ratios}
\label{app:A}

Owing to the variety of enrichment channels and the distinct timescales associated with each process, the rate of chemical enrichment varies among different chemical species. In particular, as discussed throughout this paper, iron, oxygen, and nitrogen trace the chemical evolution driven by their dominant production channels: SNIa, SNII, and AGB stars, respectively. 

\begin{figure}[h!]
    \centering
    \includegraphics[width=\linewidth]{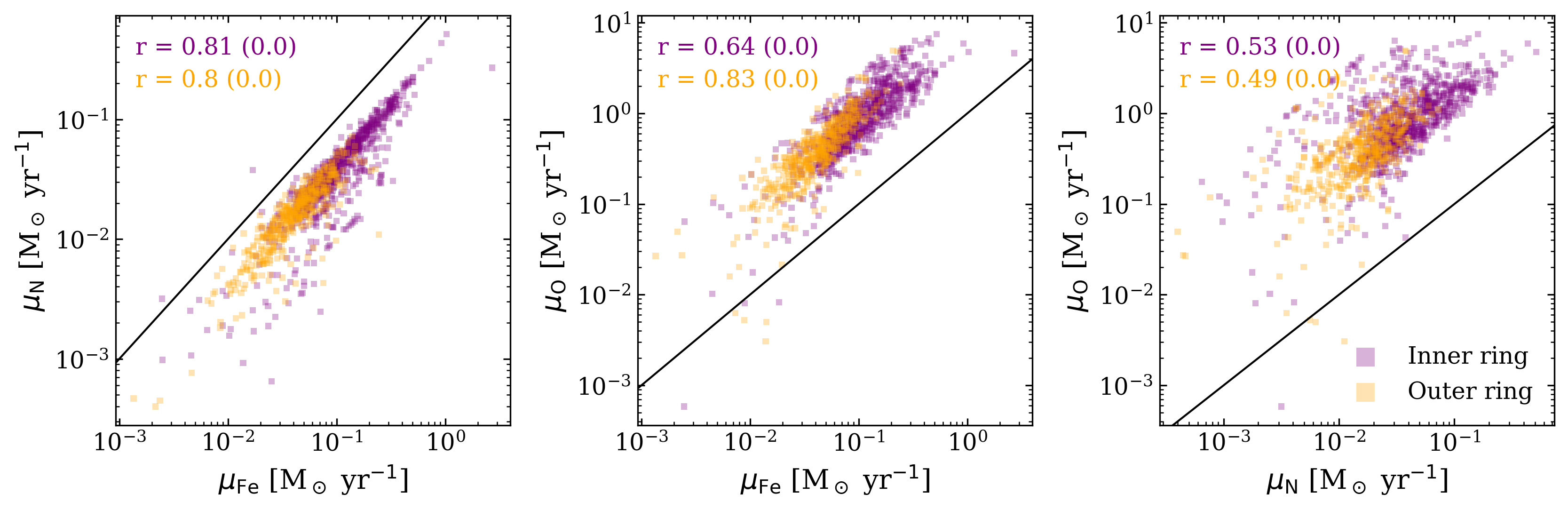}
    \caption{Chemical enrichment rates for nitrogen vs. iron (left), oxygen vs. iron (middle), and oxygen vs. nitrogen (right). The one-to-one relation is shown as a solid line. }
    \label{fig:enrichment_rates}
\end{figure}

Figure~\ref{fig:enrichment_rates} presents a comparison between the enrichment rates of different chemical species. Iron and nitrogen exhibit the tightest correlation (left panel), reflecting their similar enrichment timescales associated with SNIa and AGB stars. However, iron enrichment rates are systematically higher due to the additional contribution from SNII. The differences between inner and outer rings are significant, with inner regions typically showing higher rates—consistent with their proximity to the bulge and with the migration scenario discussed in the main body.

The middle panel shows the relation between oxygen and iron enrichment rates. While still correlated, the scatter increases compared to the nitrogen–iron relation, reflecting the shorter timescales of oxygen production dominated by SNII. The distinction between inner and outer regions is less noticeable.

The right panel compares oxygen and nitrogen enrichment rates. This relation exhibits the largest scatter among the three, consistent with the distinct stellar sources and timescales involved — SNII for oxygen and AGB stars for nitrogen.

\newpage
\section{Contribution to the $\alpha$-enrichment from different regions}
\label{app:B}

\begin{figure}[ht!]
\centering
\includegraphics[width=\linewidth]{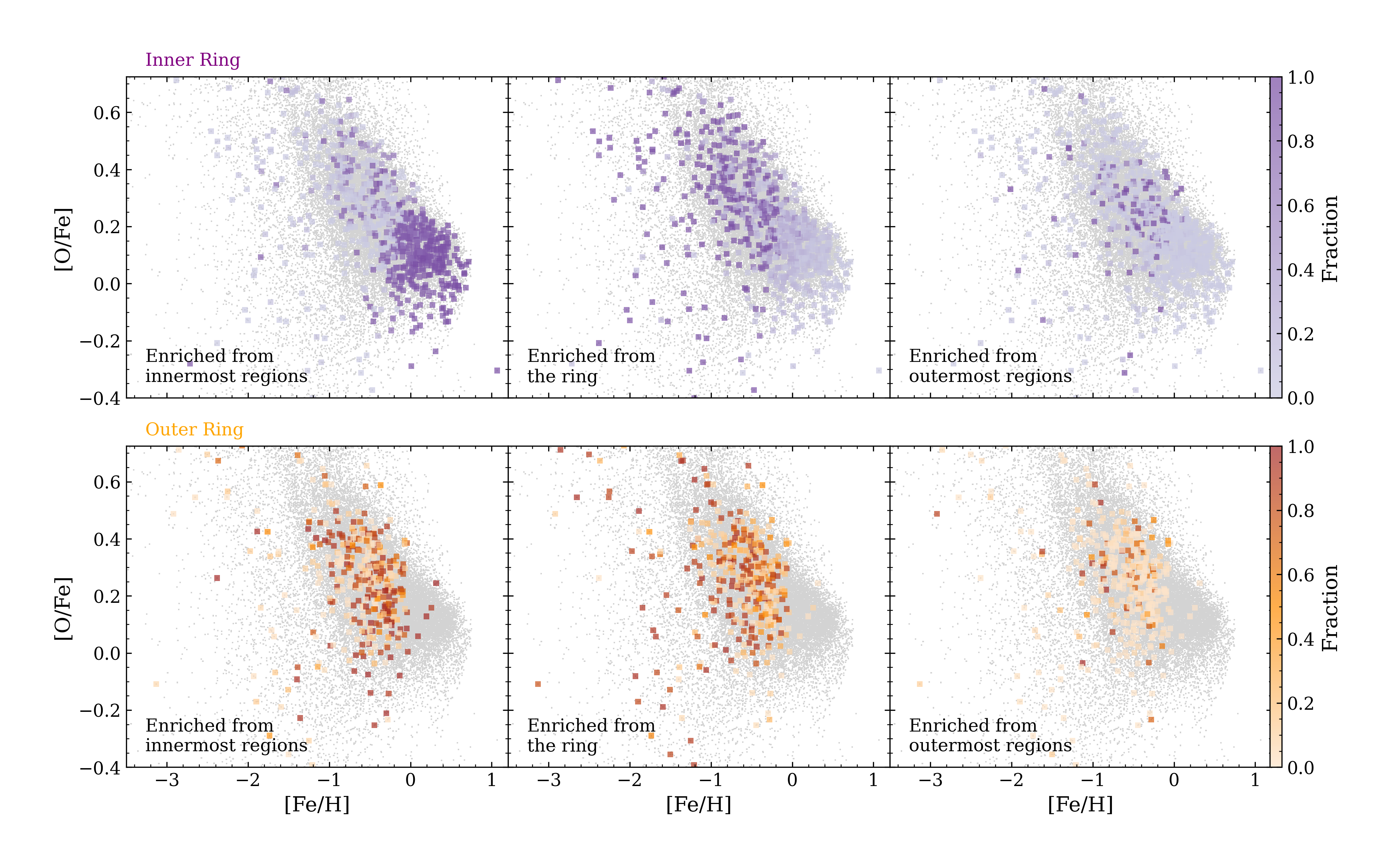}
\caption{ 
\label{fig:ofefeh}
[O/Fe]-[Fe/H] diagrams for the inner ring target stellar particles. From left to right, colour-coded by the fraction of enrichment from: (i) particles in the innermost regions; (ii) particles in the same ring; (iii) particles in the outermost regions. Darker colours indicate a higher fraction of enrichment from the respective region. We note that, for a SSP, the sum of the three contributions is always unity. 
}
\end{figure}

To visualise from where target SSPs received their chemical elements  in Fig.~\ref{fig:ofefeh}, we display the well-known $\alpha$-pattern for stars formed from gas originally in the inner and outer rings separated according to the regions where the most donors were at the time of injecting  chemical elements into the ISM. The mass fraction of contributing chemical elements is shown by the colours. 
For the stellar particles originally formed from inner ring gas, those located at the high-metallicity end of the diagram are predominantly enriched by material from the innermost region, consistent with a bar-driven gas inflow scenario. The SSPs that remain in the ring can be easily identified in the middle panel. A few particles at the low-metallicity end appear to be enriched by either the innermost or outermost regions, with a distribution that seems stochastic. In the outer ring, 
the most metal-rich and low-$\alpha$ particles are primarily enriched by material from the innermost regions, as expected. As in the inner ring case, only a small number of particles are enriched by material from the outermost regions.

\newpage
\section{Mixed inner and outer ring sample}
\label{app:C}

In this Appendix, we explore the phylogenetic tree derived from a mixed sample of stellar populations drawn from both the inner and outer rings. To ensure the same numerical representation as in the individual trees presented in Sect.~\ref{sec:phylo}, we fix the total number of SSPs to 400 by randomly selecting 200 particles from each ring. Given the extensive analysis already presented in the main text, we further ensure that the main clade identified in the inner-ring tree is fully represented in the mixed sample. To this end, all particles belonging to this clade are explicitly included in the construction of the mixed tree.

\begin{figure}[ht!]
\centering
\includegraphics[width=\linewidth]{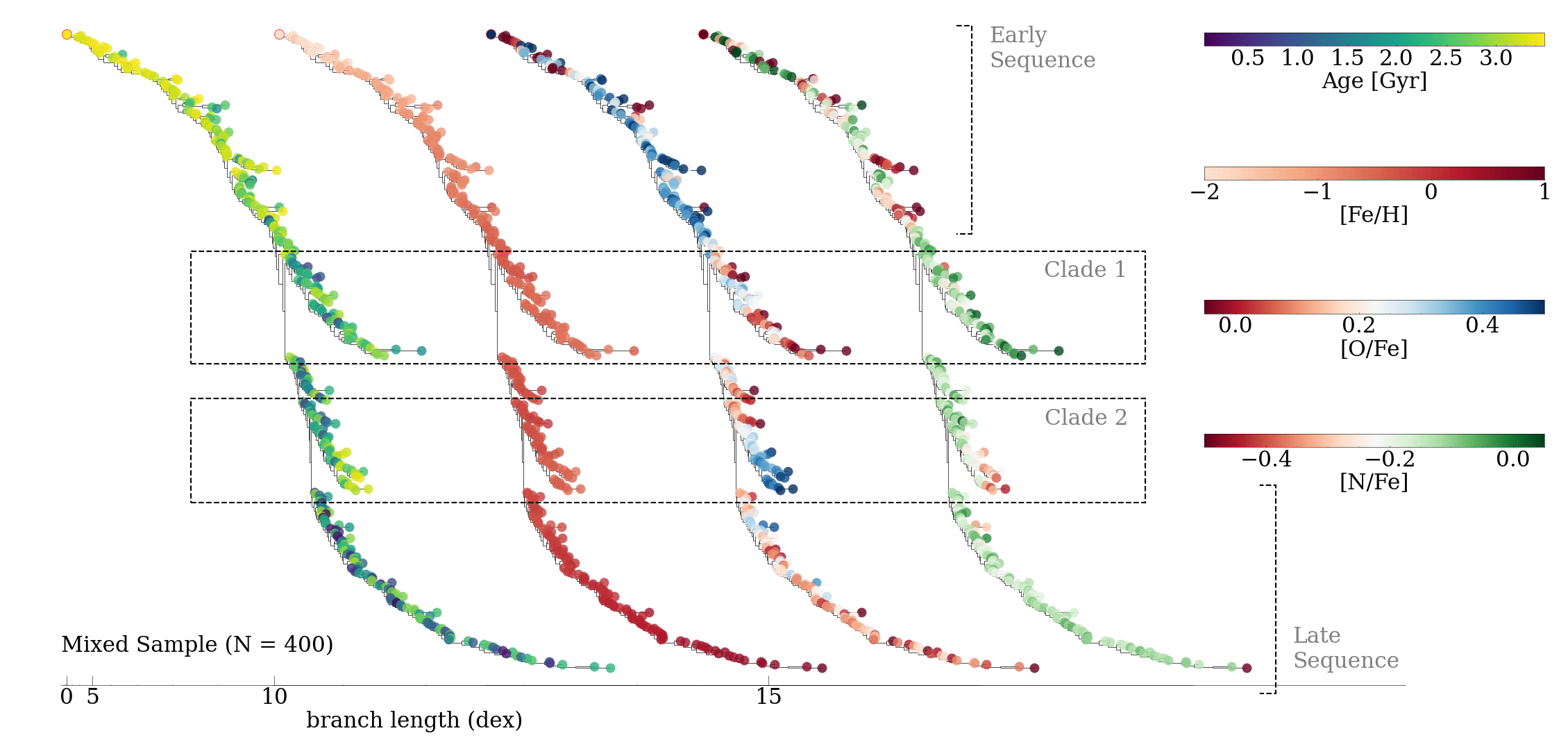}
\caption{Phylogenetic tree constructed with 400 target stellar particles selected from a mixed sample containing inner- and outer-ring particles. Colour indicates the age of the stellar particles (left), their iron abundance ([Fe/H]; centre), and their oxygen and nitrogen abundance ([O/Fe] and [N/Fe]; right).  Branch lengths are shown to scale only in the leftmost tree (using the same scaling as in Figs. \ref{fig:innertree} \& \ref{fig:outertree}); in the other panels the trees are laterally shifted for easier visual comparison.
\label{fig:appC_props}
}
\end{figure}

Figure~\ref{fig:appC_props} shows the phylogenetic tree of the mixed sample, presented in the same format as Figs. \ref{fig:innertree} and \ref{fig:outertree}. The resulting topology differs from that of the individual trees and is dominated by a prominent structure (hereafter \textit{Clade 1}). This clade is primarily composed of stellar populations with low [O/Fe] and high [N/Fe] abundances.

A secondary, smaller structure (\textit{Clade 2}) is also identified, containing stellar populations with high [O/Fe] and low [N/Fe]. In both cases, the age gradient along the branches is inverted, in qualitative agreement with the behaviour observed in the previously analysed trees.

Between the root and \textit{Clade 1}, we identify an initial sequence (hereafter \textit{early sequence}) tracing the early chemical evolution of the galaxy. Along this sequence, both age and [Fe/H] vary approximately monotonically, and most stellar populations are oxygen-rich. Variations in nitrogen abundance give rise to small substructures (<10 particles), likely associated with individual AGB enrichment events.

At the opposite end, between \textit{Clade 2} and the end of the tree, we identify a \textit{late sequence} tracing the most recent stages of chemical evolution and containing the most metal-rich particles in the simulation. This sequence is monotonic in [Fe/H], with consistently low [O/Fe] and high [N/Fe] abundances. The stellar ages follow the bifurcation of the AMR (see Fig.~\ref{fig:AMR}).

\begin{figure}[ht!]
\centering
\includegraphics[width=\linewidth]{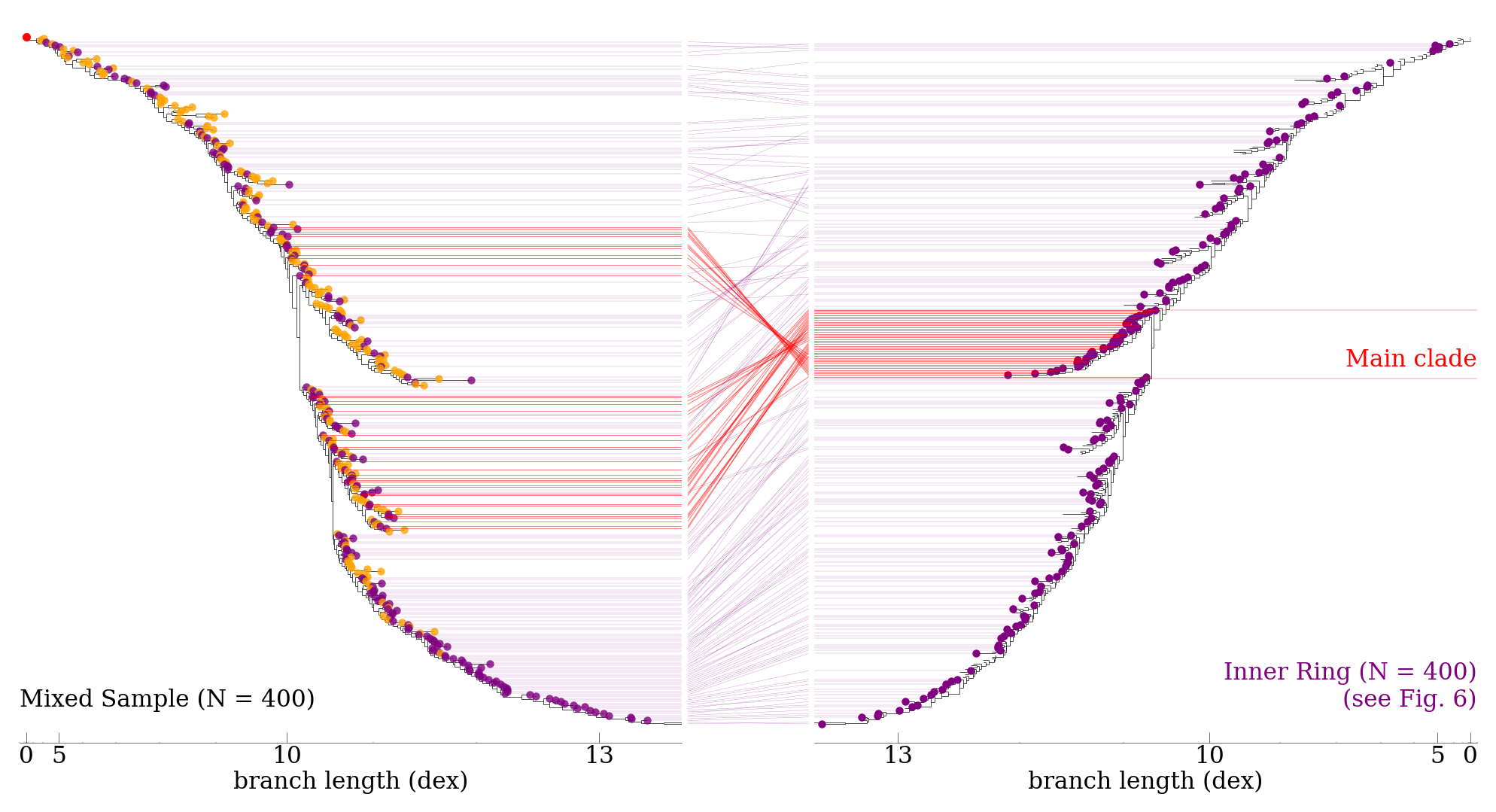}
\includegraphics[width=\linewidth]{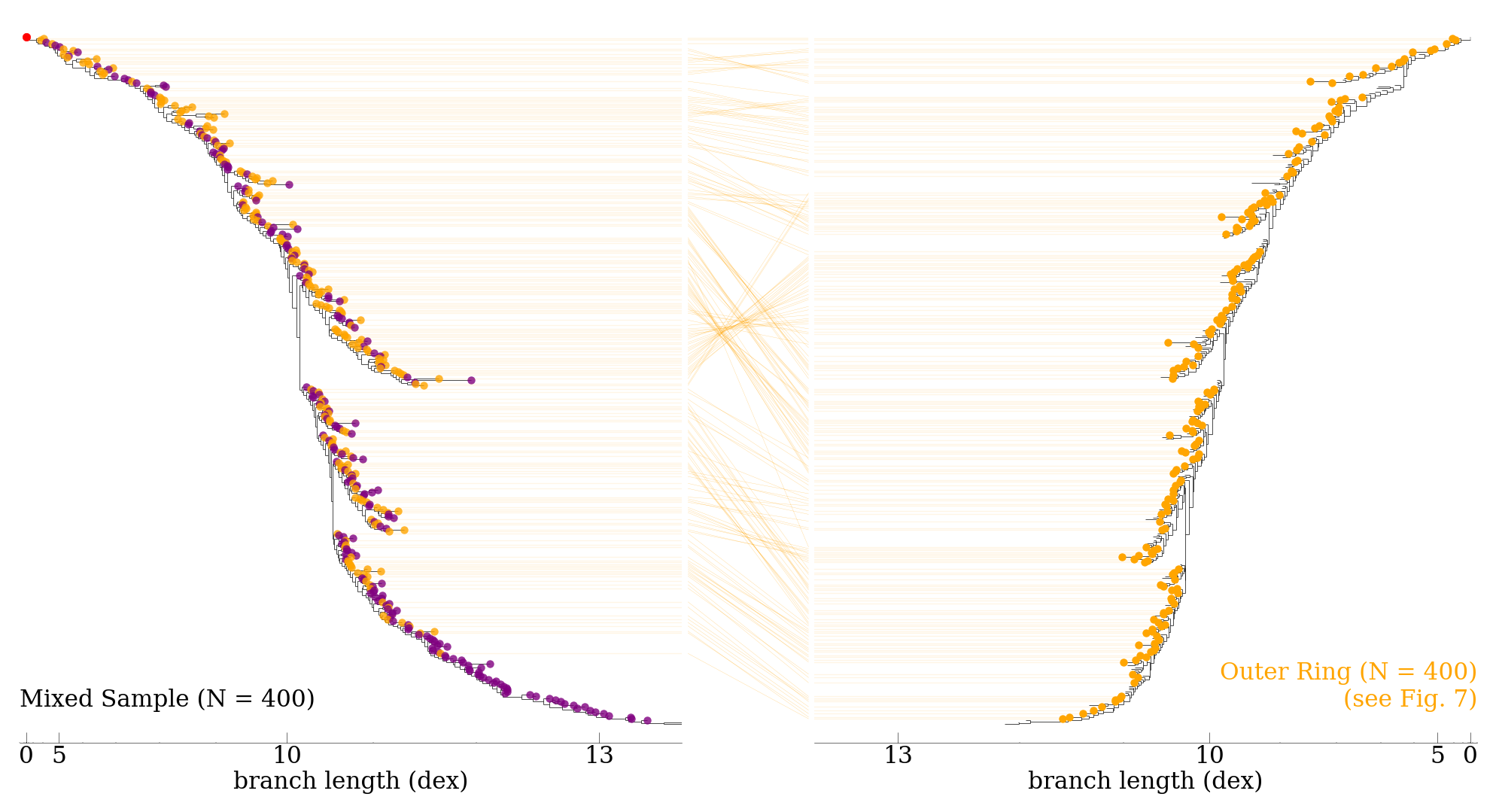}
\caption{Phylogenetic trees of the mixed sample (left panels) compared to the original inner-ring (upper right panel) and outer-ring (lower right panel) trees. Points are colour-coded according to their region of origin (purple: inner ring; orange: outer ring). In the right panels, coloured points highlight the particles selected to construct the mixed tree. Red lines indicate the members of the main clade in the original inner-ring tree. The central panels show the correspondence between the positions of the selected particles in the original trees and their rearrangement in the mixed tree.
\label{fig:appC}
}
\end{figure}

In Fig.~\ref{fig:appC}, we compare the mixed tree (left panels) with the parent trees of the inner ring (upper right panel) and the outer ring (lower right panel) analysed in the main text. Particles are colour-coded according to their region of origin (purple for the inner ring and orange for the outer ring). Horizontal lines of matching colour trace the relative positions of individual particles in the original trees and their rearrangement in the mixed tree. Red lines highlight the members of the main clade in the original inner-ring tree.

Overall, inner-ring particles tend to shift towards the tips of the tree, while outer-ring particles move closer to the root, consistent with the intrinsic chemical differences between the two samples (i.e. the outer ring hosts more metal-poor populations). Deviations from this trend are associated with the emergence of new clades or the disruption of structures present in the original trees, ultimately restoring monotonic trends with age and iron abundance. 

The \textit{early sequence} contains 62\% outer-ring and 38\% inner-ring particles. As discussed in Sect.~\ref{sec:chemicalmodel}, both rings originate from nearly pristine abundances; we therefore interpret this coexistence as evidence of a shared early enrichment history across the disc, lasting $\sim 1$ Gyr, as indicated by the youngest stellar population in the sequence.

\textit{Clade 1} is dominated by outer-ring particles (70\%), which are rearranged relative to their configuration in the original tree. This suggests that the inclusion of inner-ring populations enhances chemical contrasts, strengthening the phylogenetic signal and isolating a distinct generation of outer-ring stars.

Stellar populations in \textit{Clade 2} are nearly evenly split between the two regions (46\% inner ring and 54\% outer ring), consistent with an intermediate population sharing a similar enrichment history.

Finally, the \textit{late sequence} is predominantly composed of inner-ring SSPs (76\%), as expected from their higher metallicities. Outer-ring particles in this sequence correspond to the high-metallicity tail of their distribution. All populations in this regime are characterised by low [O/Fe] and high [N/Fe] abundances.

In summary, the mixed sample produces a phylogenetic structure that differs from those of the individual regions, while still preserving the hierarchical imprint of their distinct chemical enrichment histories. The early stages reflect a shared enrichment pathway under initially pristine conditions, whereas later structures become increasingly segregated, with \textit{Clade 1} and the \textit{late sequence} dominated by outer- and inner-ring populations, respectively, at comparable levels ($\sim 70\%$). An intermediate population emerges in \textit{Clade 2}, linking both regimes, in qualitative agreement with observational studies based on samples comprising multiple stellar populations \citep{jofre2017, jofre2025}. 

Mixing within clades highlights an important limitation of the NJ method: when inner- and outer-ring SSPs are analysed together, particles from both regions become interleaved throughout the tree and within the same clade-like structures, even when they are not linked by a direct ancestor–descendant sequence. In this sense, proximity in the tree reflects not only evolutionary history, because chemical similarity and partially shared enrichment pathways cannot be distinguished from the chemical evolution of a unique evolutionary lineage. This likely reflects the comparatively limited set of chemical traits used to characterise stellar populations, in contrast to the much higher-dimensional information encoded in DNA sequences. Nonetheless, the limitation found here using distance matrix tree building methods has been found in evolutionary biology, which has motivated sophisticated maximum likelihood tree-building techniques \citep{felsenstein2004inferring}, which consider evolutionary models as priors and expectations. Such models for galaxy evolution are work in progress.

\begin{figure}[ht!]
\centering
\includegraphics[width=0.5\linewidth]{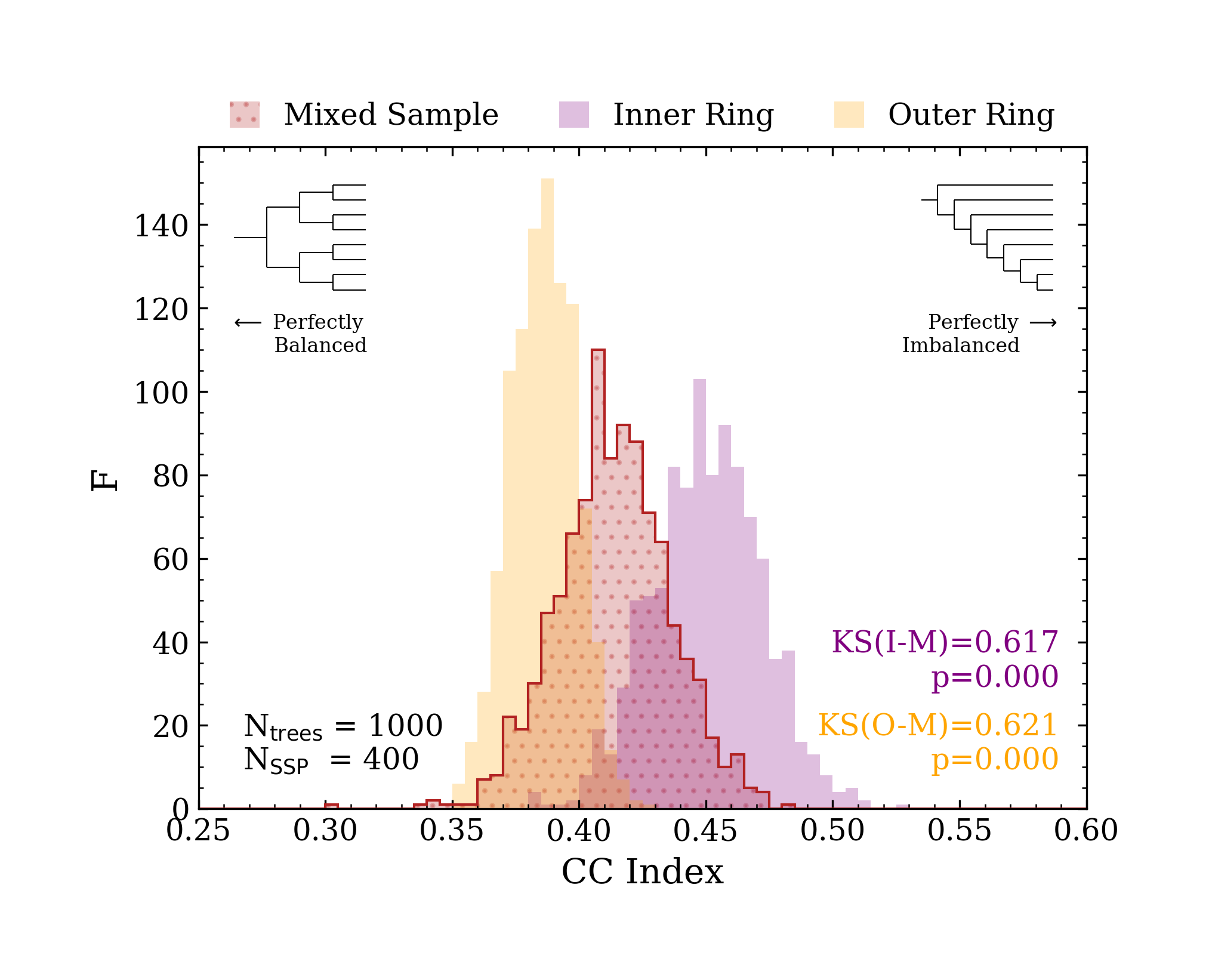}
\caption{Distribution of the CC index for 1000 phylogenetic trees built from the mixed sample (red dotted region). Distributions obtained for the inner-ring and outer-ring trees (same as Fig.~\ref{fig:indices}) are also displayed for comparison. The results of the KS tests comparing the mixed-sample distribution with those of the inner and outer rings are annotated in purple and orange, respectively.
\label{fig:appC_colless}
}
\end{figure}

To assess the statistical robustness of tree shapes, and following the methodology introduced in Sect. \ref{subsec:comparison}, we generated 1000 random realisations of mixed trees, each constructed by randomly selecting 200 inner-ring and 200 outer-ring particles. Figure~\ref{fig:appC_colless} shows the resulting distribution of the Colless imbalance index (CC Index), compared to those of the inner- and outer-ring trees presented in Fig.~\ref{fig:indices}. The mixed-sample distribution lies between those of the parent populations, indicating an intermediate degree of tree imbalance and confirming that the global phylogenetic structure reflects contributions from both components.

The tails of the distribution correspond to realisations in which the phylogenetic signal is either enhanced or diluted, depending on the specific subsample. Kolmogorov–Smirnov (KS) tests further show that the mixed-sample distribution is statistically distinct from both the inner- and outer-ring cases. Overall, these results demonstrate that the phylogenetic approach robustly captures the combined chemical evolution of multiple stellar populations, while preserving key features of their individual enrichment histories.

\end{appendix}
\end{document}